\documentclass[10pt,preprint]{aastex}

\def\kms{\,\,\,{\rm km\,s^{-1}}}                      % km / s
\def\msol{\,{\rm M}_\odot}              % M_sol = 1.989 e33 g
\def\gcc{\,\,\,{\rm g\,cm^{-3}}}

\def\ga{\,\hbox{\hbox{$ > $}\kern -0.8em \lower 1.0ex\hbox{$\sim$}}\,}
\def\la{\,\hbox{\hbox{$ < $}\kern -0.8em \lower 1.0ex\hbox{$\sim$}}\,}

\begin{document} 

%\title{Analytical theory of the stellar initial mass function.\\
%Derivation of the mass function} 

\title{Analytical theory for the initial mass function: \\
CO clumps and prestellar cores}

\author
{Patrick Hennebelle }
\affil{Laboratoire de radioastronomie, UMR CNRS 8112,\\
 \'Ecole normale sup\'erieure et Observatoire de Paris,\\
24 rue Lhomond, 75231 Paris cedex 05, France }

\and

\author{Gilles Chabrier\altaffilmark{1}}
\affil{\'Ecole normale sup\'erieure de Lyon,
CRAL, UMR CNRS 5574,\\ Universit\'e de Lyon, 69364 Lyon Cedex 07,  France}

\altaffiltext{1}{Visiting scientist, Max Planck Institute for Astrophysics, Garching, Germany}

\date{}

%%%%%%%%%%%%%%%%% END OF PREAMBLE %%%%%%%%%%%%%%%%

\begin{abstract}
We derive an analytical theory of the prestellar core initial mass function based on an extension of 
the Press-Schechter statistical formalism applied in cosmology.
Our approach relies on the
general concept of the gravo-thermal and gravo-turbulent collapse of a molecular cloud, 
with a selection criterion based
on the thermal or turbulent Jeans mass, which yields the derivation of the mass spectrum of
self-gravitating objects in a quiescent or a turbulent environment. 
With the same formalism,
simply by using a constant density threshold for the selection criterion, 
we also obtain the mass spectrum
for the non self-gravitating clumps produced in supersonic flows.
The mass spectrum of the self-gravitating cores reproduces very
well the observed initial mass function, from the high mass domain to the brown dwarf regime, and identifies the
 different underlying mechanisms
responsible for its behaviour. The theory predicts that the shape of the IMF results from two competing
contributions, namely a
power-law at large scales and an exponential cut-off (lognormal form) centered around the characteristic mass for 
gravitational collapse condition. The cut-off is not specifically due
to turbulence and exists already in the case of pure thermal collapse, provided that the
underlying density field has a lognormal distribution. Whereas pure thermal collapse produces a power-law tail
steeper than the Salpeter value, $dN/d\log M\propto M^{-x}$ with $x \simeq 1.35$, 
this latter is recovered exactly for the (3D) value of the spectral index of the velocity
 power spectrum, $n\simeq 3.8$, found in observations and in numerical simulations of isothermal supersonic
 turbulence. 
Indeed, the theory predicts that  $x=(n+1)/(2n-4)$ for self-gravitating structures and  $x=2-n'/3$ for 
non self-gravitating structures, where $n'$ is the power spectrum index  of $\log \rho$.  
%Magnetic support in our theory enters in a way similar to  the hydrodynamical contribution, simply leading to a rescaling of the Jeans mass. 
 We show that, whereas supersonic turbulence promotes the formation of
both massive stars and brown dwarfs, it has an overall negative impact on star formation, decreasing the star formation
efficiency.
This theory provides
a novel theoretical foundation to understand the origin of the  IMF and to infer its behaviour 
in different environment, 
characterized by the local properties of the gas,
from today Galactic conditions to the ones prevailing at high redshift. 
As for the Press-Schechter theory in cosmology, the present theory
provides a complementary approach and useful guidance to numerical simulations exploring star formation, while making testable predictions.
\end{abstract}


\section{Introduction}
Since the seminal work of Salpeter \cite{S55} and its empirical derivation of the stellar initial mass
function (IMF), defined as the number density of stars per log-mass interval, $dN/d\log\, M$, tremendous effort has been devoted to 
the characterization of the IMF in various environments, including the Galactic field, young clusters, star
forming regions but also the Galactic bulge and halo or high-redshift galaxies, with the aim to
identify the underlying physical mechanisms responsible for its behaviour (see e.g. Kroupa 2002, Chabrier 2003a, 2005 for recent reviews). 
A correct understanding of the IMF is one of the major, unresolved issues in astrophysics. Indeed, the
IMF provides the essential
 link between stellar and galactic evolution and determines the chemical, light and baryonic content of the universe. 
On the other hand, various observations of the prestellar dense core mass function (CMF) have shown its striking similarity with the stellar IMF, the former one being  shifted by a factor of $\simeq$ 2-4 towards higher masses
with respect to the latter one (Motte et al. 1998, Testi \& Sargent 1998, Johnstone et al. 2000, Andr\'e et al. 2007, Alfves et al. 2007). This strongly suggests 
that the IMF might already be determined by the CMF, at the prestellar stage, and thus would be due to the dynamics of the parent molecular gas. 
In the present paper, we adopt this observationally supported perspective and we refer without distinction to the CMF/IMF
for the derivation of the general theory, except in \S7.1 where we specifically address this issue.

\subsection{Previous works}
Various attempts, by many authors, have been made to
derive a  general theory for the IMF, with the aim to understand
its origin and thus to make robust {\it predictions} for the stellar mass distribution under various conditions.
These theories are based on analytical or numerical studies and, without being exhaustive,  invoke either gravitational fragmentation or 
accretion (Silk 1995, Inutsuka 2001, Basu \& Jones 2004, Bate \& Bonnell 2005), 
turbulence (Padoan et al. 1997, Padoan \& Nordlund 2002, Tilley \& Pudritz 2004, Ballesteros-Paredes et al. 2006),
 purely independent stochastic processes (Larson 1973, Zinnecker 1984, Elmegreen 1997)
or outflows (Adams \& Fatuzzo 1995) as the dominant mechanism responsible for the CMF/IMF. 

The CMF/IMF has been inferred from hydrodynamical simulations using either the SPH technique and sink particles 
(Bate \& Bonnell 2005) or grid codes and clump finding algorithms (Padoan \& Nordlund 2002, Padoan et al. 2007, Tilley \& Pudritz 2004, Li et al. 2004, Ballesteros-Paredes et al. 2006), but with different conclusions. 
Padoan et al. (2007) conclude that the CMF/IMF produced in purely hydrodynamical simulations is too stiff,
compared with the Salpeter value, whereas Tilley \& Pudritz (2004), performing a Virial analysis, reproduce
the Salpeter slope and
Ballesteros-Paredes et al. (2006) find that the CMF/IMF depends on the Mach number.
Padoan et al. (2007) also conclude that simulations including the presence of a weak magnetic field yield
 the proper  CMF/IMF and thus that the presence of a magnetic field is mandatory to recover the correct Salpeter high-mass tail.

Analytically, the most complete and successful approach to the derivation of the CMF/IMF
 has been proposed by Padoan \& Nordlund (Padoan et al. 1997, Padoan \& Nordlund 1999, 2002). It entails the following 
basic assumptions: protostellar cores result from overdensities due to shocks, and fluctuations of the underlying velocity field follow a  Kolmogorov-like scale dependence. This enables them 
to obtain the mass of the cores as a function of the scale. Then, these authors invoke the fact that the  number of cores should depend on the scale $L$
as ${\cal N}(L) \propto L^{-3}$. Combining this latter relation with their mass scale-dependence, they obtain a 
distribution of the number of objects per mass interval, $ {\cal N} = d N / d M$, which follows a powerlaw. As a last step, they combine this distribution with the 
distribution of Jeans masses obtained from the probability density function (PDF) of the field. One of their main conclusions is that 
in the hydrodynamical case, the slope of the CMF/IMF is too stiff compared with the Salpeter value, while
this latter is recovered in 
the weakly magnetized case, due to different magnetized shock jump relations. This approach,
although
retaining undoubtedly some interesting concepts, presents various shortcomings. First of all, the 
${\cal N}(L) \propto L^{-3}$ relation is not rigorously justified (see Elmegreen 2007). 
Furthermore, there is no
proper integration over a fluctuation spectrum and the Jeans mass distribution should not be independent of 
the core mass inferred from the shock conditions. The shock conditions themselves are not well justified for 
the magnetic case since they assume that the magnetic field is always perpendicular to the shock. Finally, 
the result of the CMF/IMF being too stiff in the hydrodynamical case is in contradiction with the numerical simulations 
of Bate \& Bonnell (2005) and Tilley \& Pudritz (2004). 
At last, in the Padoan \& Nordlund model, the turbulence is supposed to be taken into account
implicitly in the shock velocity dependence and through the lognormal distribution of Jeans masses,
but the turbulent support is {\it not} explicitly accounted for in this theory.

In conclusion, none of the presently available analytical theories of the mass function provides a rigorous and robust 
foundation for the CMF/IMF.

\subsection{Aim of the paper}
Such a robust theory has been derived in the cosmological context by Press and Schechter (1974, hereafter PS), who have derived an analytical formulation of the number density of
collapsing dark matter halos in the so-called hierarchical model of structure formation. In spite of its
relative simplicity, this formalism has been shown to reproduce remarkably well the results of large
computer simulations and has been applied successfully to various cosmological problems aimed
at understanding galaxy formation. This formalism also provides precious guidance to observations or numerical
simulations of structure formation in the universe. The first attempt to apply the PS formalism to the field of
star formation was made by Inutsuka (2001). In the present paper, we extend this formalism to the characterization of the CMF/IMF of stellar cores
and of CO clumps,
in the general picture of the gravo-thermal or gravo-turbulent collapse of a molecular cloud, with a selection criterion
based on the thermal or turbulent Jeans mass for the stellar CMF/IMF, and on a simple density threshold for the CO clumps. 
Our analytical theory reproduces well the characteristic
IMF derived from observations over the entire mass range from the high mass to the brown dwarf domains,
with the proper characteristic mass and Salpeter high-mass power-law exponent. 
We demonstrate that the CMF/IMF is obtained from the statistical selection involving hydrodynamical thermal or non-thermal motions,
characterized by a lognormal density distribution, and gravitational instability, through
the threshold condition for gravitational collapse, which sets up the characteristic mass.
This interplay between hydrodynamics and gravity
leads to the transition from the power-law tail at large masses to a lognormal form at small masses, as found in the IMF inferred from observations (Chabrier 2003a, 2005).
 In the presence of large-scale
turbulent motions, the slope of the power-law tail is determined by the spectral index of the turbulent power spectrum, and the proper Salpeter
exponent corresponds to the present-day observed value in molecular clouds. The characteristic mass and the variance of the lognormal part are entirely
determined by the local conditions prevailing in the cloud, namely the temperature, density and thermal or turbulent rms velocity. Although based on the same general concept as the one suggested by Padoan \& Nordlund (see references above),
the present theory is more general and identifies shortcomings in these author approach. 
The paper is organized as follows: in \S2, we derive the framework of our statistical formalism; in \S3,
we consider the simple density threshold  and calculate the related mass spectrum of CO clumps;
in \S4, we introduce
the Jeans-type selection criterion. The general CMF/IMF analytical formulation is derived in \S5, while the
results are presented in \S6. Section 7 is devoted to comparison with  observations and with 
previous works. Section 8 concludes the paper. 
In order to facilitate the reading of the paper, the most important symbols and notations used in the paper are given in Table 1.

\begin{tabular} {| {l} | {l} | }
\hline
$x$ & index of the IMF $dN/d\log M\propto M^{-x}$ (Salpeter index: $x=1.35$) \\
\hline
$n$ & 3D power spectrum index of the velocity field (Kolmogorov case: $n=11/3$)\\
\hline
$n'$ & 3D power spectrum index of the log-density ($\log \rho$) field \\
\hline
$\eta$ & exponent of the velocity dispersion vs size relation, $\eta=(n-3)/2$ \\  
\hline
$L_i$ & injection scale or system size \\
\hline
$R$ & smoothing scale  \\
\hline
$\lambda_J^0$ & Jeans length \\
\hline
$\widetilde{R}$ & $R / \lambda_J^0$  \\
\hline
$M$ & cloud mass \\
\hline
$M_J^0$ & Jeans mass \\
\hline
$\widetilde{M}$ & $M / M_J^0$  \\
\hline
$\widetilde{M}^*$ & Mass at which thermal and turbulent support are comparable \\
\hline
$\bar{\rho}$ & cloud mean mass density \\
\hline
$\delta$ & logarithmic contrast density $\log(\rho/\bar{\rho})$ \\
\hline
${\cal N}(M)$ & ${\cal N}(M) dM = dN$ number of stars of mass between $M$ and $M+dM$\\
\hline
${\cal N}_0$ & $\bar{\rho} / (M_J^0)^2$ \\
\hline
${\cal M}$ & Mach number at the cloud scale \\
\hline
${\cal M}_*$ & Mach number at the Jeans scale \\
\hline
$\sigma(R)$ & Width of the density distribution at scale $R$ \\
\hline
$\sigma_0$ & Width of the density distribution at scale $L_i$ \\
\hline
$C_s$  & sound speed \\
\hline
$V_{\rm A}$ & Alfv\'en speed  \\
\hline
$V_{\rm rms}$ & root mean square velocity, $V_{\rm rms} = V_0 (R/1 {\rm pc})^\eta$ \\
\hline
$V_{0}$ & root mean square velocity at  $R =1 {\rm pc}$ \\
\hline
\end{tabular}

\section{Statistical description} 

\subsection{PDF from supersonic turbulence}
In the PS theory of structure formation,  the structures  are identified with
over-densities in a random field of density fluctuations
\begin{eqnarray}
{\cal P}(\delta) = {1 \over \sqrt{2 \pi \sigma^2}} \exp\left(- { \delta^2 \over 2 \sigma^2} \right), 
\label{Pcosmo}
\end{eqnarray}
where $\delta = \rho / \bar{\rho}  -1 $,  ${\bar \rho}$ being the local mean density, and 
$\sigma$ is the standard deviation of the distribution. 
However, whereas the density of the primordial
universe is known to be very uniform, the density field in molecular clouds is highly non-uniform, which
considerably increases the degree of complexity to derive an analytical formalism.
On large scales ($\ga$ pc), the spectral line widths of molecular transitions observed in molecular clouds indicate 
highly supersonic motions characterized by Mach numbers ${\mathcal M}$=${\langle V^2 \rangle^{1/2} / C_s}> 1$, 
where $C_s=({kT / \mu m_H})^{1/2}\approx 0.2\,({\mu / 2.0})^{-1/2}\,({T / 10\,{\rm K}})^{1/2}\kms$ 
denotes the thermal sound speed and $\mu$ is the mean molecular weight. 

The exact nature of the turbulence in molecular clouds is still a matter of debate.
Numerically, it has been found that  non self-gravitating supersonic isothermal turbulence leads to
density fluctuations which are Gaussian at all scales in terms of the {\it logarithm}
of the density, $\delta = \log (\rho / \bar {\rho}) $ \footnote{Here we adopt a different notation 
from the cosmological case since 
$\delta$ is not the difference between the variable $\log(\rho  / \bar{\rho} )$ and its mean.}, 
yielding a so-called lognormal distribution in density
(V\'azquez-Semadeni 1994, Padoan et al. 1997,
Passot \& V\'azquez-Semadeni 1998, Ostriker et al. 2001, Kritsuk et al. 2007).
More specifically, it has been inferred from numerical simulations that
\begin{eqnarray}
{\cal P}(\delta) = {1 \over \sqrt{2 \pi \sigma_0^2}} \exp\left(- { (\delta - \bar{\delta})^2 \over 2 \sigma_0^2} \right)
\label{Pr}
\end{eqnarray}
 where $\bar{\delta}=-\sigma_0^2/2$, from eq.(\ref{Pr}), and
\begin{eqnarray}
\sigma_0^2=\ln (1 + b {\cal M}^2)
\label{sigma_val}
\end{eqnarray}
with $b\approx 0.25$. In this expression, 
the Mach number ${\cal M}$ can be either a hydrodynamical or an Alfv\'enic Mach number. 

Whether these equations accurately describe the dynamics of molecular clouds remains an open issue. 
In our approach, we will 
nevertheless adopt these expressions for reference. In any event, our formalism and our results can easily be applied to 
other distributions.

\subsection{Scale dependence of the density PDF}
\label{scale_dep}
In the cosmological literature, 
the dependence of $\sigma$ on the scale $R$ at which the density
field is smoothed has been established to be 
\begin{eqnarray}
\sigma^2(R) = \int ^{ \infty} _0 \widetilde{\delta}^2(k) W_k^2(R) d^3k, 
\label{sigma}
\end{eqnarray}
 where $W_k$ is some window function to be
 specified (see e.g. Press \& Schechter 1974, Bower 1991, Padmanabhan 1993) and 
$\widetilde{\delta}(k)$ is the {\it density} power spectrum.
This expression simply states how the variance of the density field is related 
to its  power spectrum. In the context of molecular clouds, a similar approach, 
namely the delta-variance, has been proposed by Stutzki et al. (1998) and Bensch et al. (2001).

Assuming that $\widetilde{\delta}^2$ is proportional to $k^{-n'}$ and using a window function sharply 
truncated in the $k$-space, $W_k(R) = \theta(R^{-1}-k)$, where $\theta(z)=1$ if $z>0$ and 0 otherwise,
one gets $\sigma^2(R)  \propto R^{n'-3}$. This expression is consistent provided $n'<3$, since 
otherwise the integral diverges at small $k$ (Padmanabhan 1993).

Unfortunately, in the context of supersonic turbulence, the PDF stated by eq.(\ref{Pr}) does not 
entail any scale dependence, despite the fundamental need to specify the scale at which the 
quantity is measured. This is likely due to the fact that in turbulence the fluctuations 
are dominated by the large scales and thus, as long as the measurement is done at a scale 
which is small compared to the injection length $L_i$,  the deviation from eq.(\ref{Pr})
remains small. 
In the case of weakly compressible turbulence, the density field undergoes
small fluctuations. Therefore, in that case, it is possible to refer to the cosmological 
results (Padmanabhan 1993) and to use the expression given by eq.(\ref{sigma}), as pointed out
by Hennebelle \& Audit (2007). When the turbulence is subsonic,
 the power spectrum index of the density field 
is known to be close to the Kolmogorov index obtained for the 
{\it velocity} power spectrum in incompressible turbulence, 
i.e. $n=11/3$  (e.g. Bayley et al. 1992, Kim \& Ryu 2005) 
\footnote{Note that, in the present paper, the indexes $n^\prime$ and $n$ of the power spectrum
of the log-density and velocity fields, respectively, refer to the 3D value of the power spectrum. The usual 1D
Kolmogorov and Burgers values thus correspond to $n_{1D}=n-2=5/3$ and $4-2=2$, respectively}.
In that case, however, the integral will diverge at small $k$, as mentioned above,
due to the fact that the larger scales contain most of the energy. 
Therefore, the integral must be truncated at the smallest values of  $k$, $k_{\rm min} \sim 2 \pi /L_i$,
 so that eq.(\ref{sigma}) becomes:
\begin{eqnarray}
\sigma^2(R) = \int ^\infty _{2 \pi /L_i} \widetilde{\delta}^2(k) W_k^2(R)\, d^3k
= \int ^{2 \pi/R} _{2 \pi /L_i} \widetilde{\delta}^2(k) 4 \pi k ^2 dk
 = C \left( 1 - \left( {R \over L_i} \right)^{n'-3} \right),
\label{sigma_turb}
\end{eqnarray}
where $C$ is some constant.
Note that Hennebelle \& Audit (2007) have numerically checked that this expression is reasonable.

In case of supersonic turbulence,  the power spectrum of $\log (\rho)$ has been 
calculated by Beresnyak et al. (2005) and Kritsuk (priv. comm.)
in  isothermal hydrodynamical and MHD simulations.
The derived exponent, $n'$, turns out to be 
 close to the $n=11/3$ value obtained in incompressible turbulence for the velocity field.
Note that this is not the case for the exponent of the {\it density} field power spectrum 
which becomes much smaller when the Mach number increases 
(Beresnyak et al. 2005, Kim \& Ryu 2005). 
Strictly speaking, in the case of supersonic turbulence,  the use of
eq.(\ref{sigma}) is questionable  since it is valid only for a periodic function,
which is unlikely to be the case for the density field within a spatially finite molecular cloud.  
However, more generally, $\sigma(R)$
must obey the following properties. 
First of all, $\sigma \rightarrow 0$ when $R \rightarrow L_i$;
 second of all, $\sigma$ must tend toward the expression given by eq.(\ref{sigma_turb})
when ${\cal M} \rightarrow 0$, implying $C=\sigma_0^2$; third, $\sigma \rightarrow \sigma_0$ when $R \ll L_i$. 

Since $\sigma(R)$ is rather constrained and since the expression stated by 
eq.(\ref{sigma}) with $C=\sigma_0^2$ obviously satisfies all these constraints, 
it seems reasonable to assume that
$\sigma$ is still given by eq.(\ref{sigma_turb}) even in the supersonic case. 
In any event, as will be shown later, 
 the exact dependence of $\sigma$ upon $R$ is 
not a crucial issue for the derivation of the CMF/IMF, as long as 
the aforementioned third condition is satisfied. 
Note also that our results regarding the mass spectrum of structures
defined by a density threshold  will remain
unchanged as long as $ \sigma^2(R) = \sigma_0^2 \times f ( 1 - ( R / L_i )^{n'-3})$,
with $f$ any positive and monotonically increasing function such that $f(0)=0$.

Therefore, in the following, we consider the random field of density fluctuations as given by:
\begin{eqnarray}
{\cal P}_R(\delta) = {1 \over \sqrt{2 \pi \sigma(R)^2}} 
\exp\left(- { (\delta + {\sigma(R)^2\over 2})^2 \over 2 \sigma(R)^2} \right),
\label{Pr_R}
\end{eqnarray}
where $\sigma$ is given by eq.~(\ref{sigma_turb}) with $C=\sigma_0^2$.

In our approach, the star-forming clumps issued from these large-scale turbulent 
motions are thus identified with over-densities $\delta=\log (\rho / {\bar \rho})$.
The mass associated with the surdensity at scale $R$ is
\begin{eqnarray}
M \simeq R^3 \rho =  R^3 \bar{\rho}\, e^\delta,
\label{masse}
\end{eqnarray}
so that $R = (M/\bar{\rho}) ^{1/3} \exp(-\delta/3) $. 
Note that since we use the sharp k-space filter, the relationship between $R$ and $M$ is 
ambiguous up to factor of unity.
It is important to note that, unlike in the PS formalism,
the mass depends not only on the scale $R$ but also on the variable $\delta$, 
a consequence of the lognormal, instead of uniform, underlying density distribution.

\section{Mass spectrum of structures defined by a density threshold}

Before we investigate the more complex case of self-gravitating objects, we first consider 
objects  defined as in PS by a simple  density threshold, $\rho_c$. 
In the PS formalism, the structures above this threshold  will have collapsed at the time of interest
 and will have formed a gravitationally bound object. 
In our case, this is different since in the ISM the gas is supported against gravitational collapse by 
various sources (see next section). However, structures above some density threshold may either 
undergo thermal instability, as studied by Hennebelle \& Audit (2007), or simply be revealed observationally 
because they have a sufficient abundance of CO molecules. In any case, it is interesting to 
investigate this case because of its simplicity and because there are now numerous observations of the CO mass spectrum.

%\begin{eqnarray}
%M_R= \rho_c R^3. 
%\label{MR}
%\end{eqnarray} 

The  total mass, $M_{\rm tot}(R)$, of the gas which at scale $R$ has a density larger than
$\rho_c=\bar{\rho} \exp(\delta_c)$ is simply:
\begin{eqnarray}
M_{\rm tot}(R) = L_i^3 \int ^{\delta_{sup}} _ {\delta_c} \bar{\rho} \exp(\delta) \,  {\cal P}_R(\delta)\,  d\delta,
\label{gauche1}
\end{eqnarray}
where $L_i$ is the size of the system  assumed to be comparable to the injection scale.
The value $\delta_{sup}$ is not consequential on the results and  will be
assumed to be equal to infinity, $\delta_{sup}\rightarrow \infty$, in the rest of the calculations.
This total mass, $M_{\rm tot}(R)$, represents the mass of the structures of size larger than or equal to $R$ 
and of mass larger than or equal to 
\begin{eqnarray}
M^c_R= \rho_c R^3. 
\label{MR}
\end{eqnarray} 

We are interested in counting the structures having a mass equal to $M_R^c$.
Some of these structures, however, have a non-zero probability to be included in larger structures which  
exceed the density criterion, and thus the number of these structures cannot be straightforwardly
obtained by requiring that the total mass they contain be equal to $M_{\rm tot}$. 
This is similar to the so-called cloud in cloud problem in the PS theory \cite{Bond91, Jedam}. 
We follow the approach of Jedamzik (1995) to handle this problem (see also Yano et al. 1996 and Nagashima 2001). 

Let $ {\cal N} (M')  dM'$ be the number-density of isolated structures of  mass
between $M'$  and $M'+dM'$, which  satisfy the aforementioned density criterion.
More precisely, these structures are the ones which at scale $R' \simeq (M' / \rho_c)^{1/3}$
would have a density $\bar{\rho} \simeq \rho_c$.  
The mass contained in  such structures is $M' {\cal N}(M') dM'$. 
At scale $R < R'$, these structures contain regions whose density can be 
larger or smaller than $\rho_c$.
%Let $P(M,M')$ be the probability of finding a structure of  mass 
%$M$ satisfying the criterion and embedded inside a  structure of  mass $M'$.

Let $P(M_R^c,M')$ be  the  mass fraction of a  structure of mass $M'$ which, at scale 
$R$, has a density above the critical density.    
The mass of the gas having a density larger than $\rho_c$ at scale $R$ and yet 
contained in a structure of mass $M'$ as defined above is thus:  $M' {\cal N}(M') P(M_R^c,M')  dM'$.
This yields a second expression for $M _{\rm tot}(R) $:
\begin{eqnarray}
M_{\rm tot}(R) = L_i^3\, \int _{M^c_R}  ^\infty  M'\,   {\cal N} (M')\,   P(M^c_R,M')\,  dM'.
\label{droit1}
\end{eqnarray}
 As emphasized by Jedamzik (1995), the exact value of $P(M,M')$ 
depends on the choice of the window function and to some extent on the exact definition of a structure
(see Yano et al. 1996). Jedamzik explored the 
influence of  $P(M,M')$  and found that its value has some limited influence on the result. 
However, in the case of a window function sharply truncated 
in the k-space, Yano et al. (1996) argue that $P(M,M')=1/2$.
This stems from the fact that, in that case, the smaller region inside the larger region has an equal
probability to be overdense or underdense than $\delta_c$. 
Note that in the 
cosmological case, the Bond et al. (1991) result is exactly recovered if one assumes $P(M,M')=1/2$.
Since, as demonstrated in Appendix~\ref{p=1/2}, it appears that the calculation of Yano et al. (1996)
 is also applicable in our case, we   take $P(M_R^c,M')=1/2$.

Equating eqs.~(\ref{gauche1}) and~(\ref{droit1}) and deriving the expression with respect to $R$, 
we obtain:
\begin{eqnarray}
 {\cal N}(M^c_R)  =
 -{2 \bar{\rho} \over M^c_R} {d R \over d M^c_R} \left( \int _  {\delta_c}^\infty \exp(\delta)\, {d {\cal P}_R \over dR}\, d\delta \right)
\label{nclump}
\end{eqnarray}
This yields the following expression for the seeked mass distribution
at scale $R$:
\begin{eqnarray}
{\cal N}(M^c_R) = -{ 2 \bar{\rho}  \over  M^c_R  } {d R \over dM^c_R }  
{ 1 \over \sqrt{2 \pi} \sigma^2} {d \sigma \over dR}
\int _{\delta_c} ^\infty A(\delta,R)\, d\delta,  
\label{form_n}
\end{eqnarray}
where 
%\left( { R \over L_i} \right)^{(n-3)} 
%{ b\, (n-3) {\cal M}_0^2 \over  \sqrt{2 \pi}  \sigma ^3 (1 + b {\cal M}^2)}
\begin{eqnarray}
A(\delta,R) = \left( -1 + {\delta^2 \over \sigma^2} -{\sigma^2 \over 4} \right)  
\exp\left( -{\delta^2 \over 2 \sigma^2}  + {\delta\over 2} - {\sigma^2 \over 8 } \right), 
\end{eqnarray}
and, with the expression adopted in \S\ref{scale_dep},  
\begin{eqnarray}
{d \sigma \over d R} = -{n'-3 \over 2 \sigma} {\sigma_0^2 \over R } \left( { R \over L_i} \right)^{n'-3}.  
\label{dsig}
\end{eqnarray}
The integral can be calculated analytically:
\begin{eqnarray}
\int _{\delta_c} ^\infty A(\delta,R) d\delta = \left( \delta_c + {\sigma^2 \over 2} \right) 
\exp \left(  -{(\delta_c-{\sigma^2\over 2} )^2   \over 2 \sigma^2}  \right) 
\label{form_A}
\end{eqnarray}
and is positive if $\delta_c > - \sigma^2/2=\bar{\delta}$, as in PS. 
As demonstrated in the appendix~\ref{voids}, when $\delta_c < - \sigma^2/2$, which physically corresponds to 
a void instead of a cloud, $\delta_c + \sigma^2/2$
 must be replaced by $-(\delta_c +  \sigma^2/2)$ in eq.(\ref{form_A}), therefore 
insuring the positivity of ${\cal N}(M)$. 

With the aforementioned definition of $M_R^c$ (eq.\ref{MR}), we get for the mass spectrum of structures
defined by the density threshold $\delta_c$:
\begin{eqnarray}
{\cal N}(M^c_R) = { \bar{\rho}  \over (M^c_R) ^2 } {  (n'-3) \sigma_0^2 \over 3 \sqrt{2 \pi}  \sigma ^3  } \left( { M^c_R \over M_{0}} \right)^{{n'-3\over3} }
\left( {\bar{\rho}  \over \rho_c } \right)^{{n'-3\over3}} 
 \times 
%\nonumber
\left( \delta_c + {\sigma^2 \over 2} \right) 
\exp \left(  -{(\delta_c-{\sigma^2\over 2} )^2   \over 2 \sigma^2}  \right),  
\label{spec}
\end{eqnarray}
where $M_0=\bar{\rho} L_i^3$ is the whole mass contained within a volume
$L_i^3$ and $\rho_c = \bar{\rho} \exp(\delta_c)$. 
This expression is very similar to the one derived by PS in the
context of a uniform density background. As in PS, we identify a {\it power-law contribution} and a
 {\it Gaussian truncation} around the
threshold $\delta_c$, and we recover the fact that the
fraction of bound objects of mass greater than $M_R$ is proportional to $\delta_c -\bar{\delta}$.
When the scale $R \rightarrow L_i$, then $\sigma \rightarrow 0$ and thus there is also  
a Gaussian cut-off for large-mass structures. Note that in the limit when $\delta_c \rightarrow 
\bar{\delta} = - \sigma^2/2$, ${\cal N}(M) \rightarrow 0$ except if $\sigma \rightarrow 0$ which precisely occurs
when $R \rightarrow L_i$. This implies that, when the threshold density
$\delta_c \rightarrow \bar{\delta}$, all the mass lies within a unique structure of size $L_i$, as expected.

On the other hand, there is no cut-off for small mass
structures since they arise because of turbulent fluctuations which are scale free. This remains valid 
as long as 
the turbulent cascade remains self-similar, which implies that the scale $R$ must be large with respect 
to the dissipation scale.  
Another important difference with the PS formalism is that time dependence is not taken into account. This point
will be discussed in \S~\ref{time dep} for the case of self-gravitating fluctuations. Here, we note that 
in the context of non self-gravitating structures defined by a  simple density threshold, 
no time evolution is required to compute the mass spectrum of these structures.

Equation (\ref{spec}) naturally yields the scaling relation ${\cal N}(M) \propto { M^\beta}$ with $\beta={-2+(n'-3)/3}$, 
so the power-law exponent is now affected by the spectral index characteristic of the  logarithmic density
 power spectrum. This exponent is identical to the exponent obtained  analytically by Hennebelle \& Audit (2007, their eq.(15)) for 
subsonic turbulence and is in good agreement with the mass spectrum inferred from numerical simulations 
(Hennebelle \& Audit 2007, Hennebelle et al. 2008, Heitsch et al. 2008).
The exponent of the logarithmic density  power spectrum has received only little attention 
but, as mentioned previously, values around the Kolmogorov exponent, $n=11/3$, have been inferred from
numerical simulations (Berezniak \& Lazarian 2005, Kritsuk, priv. comm.). 
For $n'=11/3$, we get $\beta=-16/9=-1.777$.
This value is   remarkably close to the slope of the mass spectrum inferred for the CO clumps 
(Blitz 1993, Heithausen et al. 1998, Kramer et al. 1998). This indicates that the CO clumps very 
likely have a turbulent origin. Note that in Hennebelle \& Audit (2007), it was proposed
that fluctuations of the warm neutral medium (WNM) induced by a sub-transonic turbulence are amplified by
 thermal instability once the density reaches the 
instability threshold. The present calculations indicate that the two mechanisms, namely subsonic turbulence 
followed by thermal instability,
and supersonic turbulence lead to similar mass spectra. It is therefore difficult from observations of 
the mass spectrum only to 
determine which process is dominant. Note that the two mechanisms are actually not exclusive from each other.

\section{Selection criterion. Thermal and turbulent Jeans mass}
Unlike in the cosmological case, where the gas is very cold and not turbulent, 
the gas in the interstellar medium is supported by a combination of thermal pressure, turbulence and 
magnetic field. In this section we examine these various supports and we derive criteria to be fulfilled by the structures in order to collapse.

\subsection{Thermal Jeans mass}
In order for a  cloud  to collapse, its mass must be larger than the thermal Jeans mass: 
\begin{eqnarray}
M_J = a_J {C_s^3 \over \sqrt{G^3 \rho} }= a_J {C_s^3 \over  \sqrt{G^3 \bar{\rho}} }\exp \left(-{\delta \over 2} \right)=M_J^0\exp 
\left(-{\delta \over 2} \right),
\end{eqnarray}
where 
\begin{eqnarray}
M_J^0= a_J\,{ C_s^3 \over \sqrt{G^3 \bar{\rho}}}\approx 1.0\,\, a_J \,({T \over 10\,{\rm K}})^{3/2}\,
({\mu \over 2.33})^{-1/2}\,({{\bar n} \over 10^4\,{\rm cm}^{-3}})^{-1/2}\, \msol
\label{jeans}
\end{eqnarray}
where $a_J$ is a dimensionless parameter of order unity 
which takes into account the geometrical factor. 
In the absence of turbulent support, the clumps with mass larger than a Jeans mass will eventually collapse and form gravitationally bound objects. 
This implies a lower limit on the local density fluctuation,
$M \ge M_J = M_J^0 \exp(-\delta/2) \Rightarrow \delta \ge - 2 \ln \left( M / M_J^0  \right)$. 

Therefore, using eq.(\ref{masse}), we obtain two equivalent  conditions for the star forming collapsing structures: 
\begin{eqnarray} 
M \ge M_R^c = a _J ^{2/3} \, {C_s^2 \over G} R  
{\hskip 1.cm}{\rm and}	{\hskip 1.cm} 
\delta \ge \delta_R^c = -2 \ln \left( {R \over \lambda_J^0} \right)
\label{crit_y}
\end{eqnarray} 
where $\lambda_J^0= a_J^{1/3}C_s/\sqrt {G{\bar \rho}}\approx 0.1\, a_J^{1/3} \,(T / 10\,{\rm K})^{1/2}\,
(\mu / 2.33)^{-1/2}\,({\bar n} 
/ 10^4\,{\rm cm}^{-3})^{-1/2}$ pc is the thermal Jeans length \footnote{Strictly speaking the 
thermal Jeans length is $ \sqrt{\pi} C_s / \sqrt{G{\bar \rho}}$}.

An important difference with the case of \S3 is that the threshold now depends on the scale $R$. 
In particular $\delta_R^c \rightarrow \infty$ as $R \rightarrow 0$.
Physically, this means that it is more difficult to have a gravitationally unstable object at small scales
because of the thermal support. 

\subsection{Turbulent Jeans mass}

If turbulence is significant, the turbulent support must be taken into account. 
As Tilley \& Pudritz (2004), we use the Virial theorem to decide whether or not a structure is going to collapse.
Neglecting the surface terms, the Virial theorem can be written as:
\begin{eqnarray}
{1 \over 2} { d^2 I \over dt^2 } \simeq 2 E_{\rm cin} + E_{\rm pot} + (3 P_{\rm th} + E_{\rm mag}) {\cal V}
\label{viriel}
\end{eqnarray}
where ${\cal V}$ is the volume  and $E_{\rm cin}$ includes the contribution from turbulence, $\sim 1/2\,\rho \langle V_{\rm rms}^2\rangle$, 
where $\langle V_{\rm rms}^2\rangle ^{1/2}$
is the turbulent rms velocity. $E_{\rm pot}$ is the gravitational energy and $E_{\rm mag}$ the magnetic energy. 
It is usually admitted that the structures which collapse are the ones having $d^2 I / dt^2  < 0$,
although this criterion is not entirely rigorous, in particular if we were to apply it to one specific object.
 It is, however, a reasonable approximation for a statistical approach over a large 
population of objects, as in the present context (see e.g. Dib et al. 2007). 
This leads to 
\begin{eqnarray}
  \langle V_{\rm rms}^2\rangle    + 3\, (C_s)^2 < - E_{\rm pot} / M
\label{viriel_ceff}
\end{eqnarray}
Therefore, within our analytical
formulation and our statistical description of the IMF, turbulent support can be included under the usual form of an
effective sound speed (or equivalently effective pressure) (Chandrasekhar 1951, Bonazzola et al. 1987,
V\'azquez-Semadeni \& Gazol 1995)\footnote{Note that, because of the anisotropic nature of turbulent support, the effective
sound speed formulation should represent an upper limit of the true non-thermal contribution.}, 
\begin{eqnarray}
C_{\rm{s,eff}}  = [(C_s)^2 + {1\over 3} \langle V_{\rm rms}^2\rangle ]^{1/2}.
\end{eqnarray}

The turbulent rms velocity is observed to follow a power-law correlation with the size of the
region, the so-called Larson-type relations
\begin{eqnarray}
\langle V_{\rm rms}^2\rangle =  V_0^2 \times \left( {R \over  1 {\rm pc}} \right) ^{2 \eta}
\label{larson}
 \end{eqnarray}
with $V_0\simeq 1 \kms$ and $\eta \simeq 0.4$-0.5 \cite{Larson81}. 
Note that, strictly speaking, the Larson relations are representative of the molecular gas at the
whole cloud scale.
Given the low efficiency of star formation and the relatively high densities at which star formation
is observed to take place, ${\bar n} \sim 10^4$-$10^5 \gcc$ (Motte et al. 1998, Andr\'e et al. 2007), 
it is not clear whether the Larson relations are representative of star formation regions. Indeed, 
the observed line-width vs size relation for prestellar massive cores in star forming regions show 
slightly higher densities and velocity dispersions than predicted by the Larson relations 
(Caselli \& Myers 1995). Therefore, the use of the Larson relations, although reasonable, should be 
considered as simply indicative. 

Using the fact that $\langle V_{rms}^2(R)\rangle=\int_{2\pi/R}^\infty P_V(k)d{\vec k}\propto R^{n-3}$, where $P_V(k)\propto k^{-n}$ is the velocity power spectrum, and eq.(\ref{larson}), the exponent
$\eta$ is related to the (3D) index of turbulence $n$  by:

\begin{eqnarray}
 \eta=\frac{n-3}{2}.
\label{eta}
\end{eqnarray} 
Note that, strickly speaking,
the index of turbulence $n$ which appears in this expression, which is related to the power spectrum of $v$,
is not necessarily the same as the one 
previously introduced, $n'$, which is related to the less standard power spectrum of $\log \rho$.  
However, as mentioned earlier, numerical simulations seem to find that both indexes are rather similar. 
 
The aforementioned values of $\eta$ thus lie between the ones corresponding to a Kolmogorov, $\eta =1/3$, and a Burgers, $\eta =1/2$, value,
pointing to mildly to highly
supersonic conditions in star forming clouds. 
Recent high resolution simulations of non-magnetized isothermal supersonic turbulence (Kritsuk et al. 2007)
 yield $\eta \sim 0.4$-0.45 ($n \sim 3.8$-3.9).
In the limit where thermal support can be neglected (valid for massive stars), 
we obtain a turbulent Jeans mass
\begin{eqnarray}
M_{J,{\rm turb}} = a _J {V_0^3 \over  \sqrt{  3^3 G^3 \bar{\rho} \exp(\delta) } }  
\left( {R \over 1 {\rm pc}} \right)^{3 \eta}.  
\end{eqnarray}
As for the pure thermal case, the structures which collapse in the turbulent case are the ones such that $M > M_{J,{\rm turb}}$,
which now yields:
\begin{eqnarray} 
M \ge M_R^c =  a _J^{2/3} {V_0^2  \over 3 G} { \left({R \over 1 {\rm pc}} \right)^{2 \eta}  } R   {\hskip 1.cm}{\rm and}{\hskip 1.cm} 
\delta \ge \delta_R^c = \ln \left[{ a_J^{2/3} \over 3} {  V_0^2 \over G \bar{\rho} R^2 \,}  ({R \over 1 {\rm pc}})^{2 \eta}\right].
\label{crit_yt}
%\label{crit_M}
\end{eqnarray}

\subsection{General case}
\label{gen case}

In the general case where both thermal and turbulent supports contribute, 
the condition for collapse, $M > M_J$, becomes
\begin{eqnarray}
M > a_J { \Bigl[ (C_s)^2 + (V_0^2 / 3) (R /   1 {\rm pc} )^{2 \eta} \Bigr]^{3/2}   \over \sqrt{G^3 \bar{\rho} \exp(\delta) }  },
\label{cond_tot}
\end{eqnarray}
which, with eq.(\ref{masse}), implies
\begin{eqnarray} 
M > M_R^c = a_J^{2/3} 
\left( {  (C_s)^2 \over G    } R + {V_0^2  \over 3\, G  } \left({R \over 1 {\rm pc}}\right)^{2\eta} R \right)  
\label{crit_Mtot}
\end{eqnarray}  
and
\begin{eqnarray} 
\delta > \delta_R^c = \ln \left( { a_J^{2/3} [ (C_s)^2 +  (V_0^2 / 3) \left(R / 1 {\rm pc} \right)^{2 \eta} ] \over G \bar{\rho} R^2    }\right).
\label{crit_ytot}
\end{eqnarray} 

%For the case $\eta=0.5$, $R$ and thus $\delta_R^c$ can be expressed trivially as a function of $M_R^c$, with
%\begin{eqnarray}
%R = { 3 G L_i \over 2 V_0^2} \left( -{(C_s)^2 \over G}    + 
%\sqrt{ { (C_s)^4\over G^2}  + {4 \over 3} {V_0^2 \over G L_i} M_R^c} \right) 
%\label{crit_Rtot}
%\end{eqnarray}

An important point to be stressed is that, in our approach, we select the regions of the gas at the
very early stages of star formation, which will collapse {\it in the future} because
the combination of all supports is not sufficient to balance gravity. In particular, the dense cores themselves,
which are observed not to be very turbulent, represent in our approach the (collapsing) evolution from a selected initially more dilute and turbulent piece of
fluid. This piece of fluid, which fulfills our criteria for collapse, is properly accounted for in the present theory, while the observed prestellar cores
represent a more evolved state, of which physical properties are not described in the present theory. Similarly, it is known that turbulence dissipates in about one 
crossing time, so one may worry about not taking this effect into account in the theory. This can be understood as follows: if 
a piece of fluid contains too much turbulence, it will expand  and gravity will be unable to take over. On the other 
hand, if the turbulence is not sufficient, gravitational collapse will proceed eventually. The decay of turbulence does not affect 
this selection process, unless perhaps for the cases where turbulence is just sufficient to balance gravity.

\subsection{Magnetic field}
The presence of a magnetic field, known to be dynamically significant in star formation (Crutcher 1999), will modify to some extent this general picture of the origin of the CMF/IMF.
Taking into account the magnetic field in such a theory requires to know exactly how magnetic field 
and density correlate. In numerical simulations, a broad correlation has been found between density and magnetic field, 
yielding $B \propto \sqrt{\rho}$ (Passot et al. 1995, Padoan \& Nordlund 1999). Observationally,
using Crutcher's (1999) data, Basu (2000) has shown that the magnetic intensity follows very closely 
the relation $B \propto   \rho ^{1/2} \Delta V$,  where $\Delta V\equiv \langle V_{\rm rms}^2\rangle^{1/2}$ is the velocity dispersion.
This suggests that the Alfv\'enic Mach number, ${\mathcal M_A}=\Delta V / V_A$, where $V_A= B / (4 \pi \rho) ^{1/2}$ is
 the Alfv\'en
velocity, may be approximately constant in molecular clouds.
Basu proposes an explanation based on magnetic flux and mass conservation and approximate mechanical equilibrium along the field lines.  His calculation shows that the Alfv\'en
velocity can be 
written as
$V_A = (V_A^0/C_s) \times   \sqrt{(C_s)^2 + V_{\rm rms}^2/ 3} $.
Application of the Virial theorem (eq.\ref{viriel}) yields
\begin{eqnarray}
  3\,C_s^2 + {\langle V_{\rm rms}^2\rangle }    + {\langle V_A^2 \rangle \over 2}     < - E_{\rm pot} / M.
\label{viriel_mag}
\end{eqnarray}

Therefore, in order to take into account the magnetic support in expression~(\ref{cond_tot}), we just
have to replace 
$[(C_s)^2 + V_{\rm rms}^2 /3]$ by  $[(C_s)^2 + V_{\rm rms}^2 /3 + V_A^2/6]$. Assuming the
aforementioned dependence for the magnetic field, inclusion of magnetic support in our theory thus
simply leads to an expression for the Jeans length and
the Jeans mass which is 
proportional to the one derived in the hydrodynamical case. All the calculations conducted in this paper
 could thus, in principle, be generalized to the MHD case by, e.g., simply replacing $a_J$ by $a_J \times [1+ (V_A^0/C_s)^2 / 6 ]^{3/2}$
or, more generally, by simply rescaling the sound speed and the rms velocity by different factors.

The magnetic field also modifies the value of the Mach number which appears in the width of the lognormal distribution. 
This implies that, in this simple approach, the magnetic field,
because of the magnetic support,  reduces the width of  the density PDF (V\'azquez-Semadeni et al. 2005).  
Although a  more realistic treatment in which the  magnetic field distribution would be properly taken into account 
is strongly needed, it seems difficult to perform it at this stage. Meanwhile, the present approach has the virtue of simplicity,
while based on observational arguments. %More theoretical knowledge is required.  

\subsection{Time dependence issue}
\label{time dep}
In the PS formalism, time dependence is accounted for by relating the collapse epoch of the perturbation 
to its density contrast. Although this somehow simplistic approach can be improved (see e.g. Audit et al. 1997),
it  nevertheless yields satisfying results (Efstathiou et al. 1988, Lacey \& Cole 1994). 

In principle, such a time dependence could also be obtained in our case by selecting at a time $t$ not
the gravitationally unstable pieces of fluid, but rather the ones that had time to collapse and 
form a singularity. However, 
the time evolution is more complex in molecular clouds than in the cosmological case,
 because of the various sources of support of the gas
against gravitational collapse. Therefore, deriving the time at which collapse really occurs is a complicated
task. Ignoring time dependence in our theory implies that the distribution of collapsing objects obtained at the very early stages
by our threshold conditions is the one obtained once star formation 
is completed within the cloud of interest. 

A related problem concerns the dense core formation, subsequent accretion and possible merging 
(see Dib et al. 2007 and Peretto et al. 2007)  not taken into account 
in a time independent analysis. As emphasized in \S~\ref{gen case}, the present analysis relies 
on a simple {\it statistical 
counting} of the smallest  pieces of fluid which are dominated by gravity. Dense cores will form out 
of these pieces by progressively accreting the related initial reservoir of mass. 
Our analysis excludes any external accretion
of gas which is not included in this initial reservoir. 

This lack of time dependence is certainly a limitation of the present theory. However, the agreement between this
theory and the observed CMF/IMF, as shown below, seems to suggest that time evolution should not drastically affect the initial mass spectrum of collapsing structures.

\section{Mass spectrum of self-gravitating objects:  derivation of the CMF/IMF}

\subsection{Analytic formulation}

\subsubsection{Physical assumptions}
Let us consider a region of scale $R$. The places where $\delta$ is larger
than the aforederived critical threshold $\delta_R^c$,
as defined by eqs.(\ref{crit_y}), ~(\ref{crit_yt}) and~(\ref{crit_ytot}), 
contain more than one Jeans mass  and are going to
form stars of mass {\it smaller} than or equal to $M_R^c$. 

This is because we assume that the final mass of the cores which form 
is equal to the mass associated with the clouds containing only one Jeans mass,  since
clouds which contain initially more than one Jeans mass are likely to fragment into several 
objects, whose number is more or less equal to the number of Jeans masses contained in the cloud.
Therefore, all points which, at scale $R$, have a density contrast larger than $ \delta_R^c$
are going to form  structures of mass {\it smaller} than or equal to  $M_R^c$.
This can also be understood in the following way. Consider a cloud which at scale $R$ contains
about one Jeans mass, implying that its mean density at scale $R$ is about 
$M_J/R^3$. At smaller scales, it may happen that the cloud is not uniform but is composed of  smaller, 
 denser cores embedded into a more diffuse envelope. If these denser cores contain one Jeans mass,
the end product of the collapse is likely to be a cluster of objects whose mass is close to the mass of the smaller/denser cores and not to the 
mass of the object at scale $R$. We note here a fundamental difference with the cosmological case, where it is assumed 
that the mass of the final objects is equal to the mass of the biggest cloud which satisfies the appropriate 
conditions. This difference arises from the fact that the density contrast between the standard molecular gas and the 
star itself is about 18-20 orders of magnitude. Note that in our approach, as mentioned earlier, we neglect any further accretion
on the prestellar cores. More precisely, it is assumed that most of the mass which will eventually
be accreted onto the prestellar dense core is contained within the initial Jeans mass reservoir.

Given the aforederived threshold conditions for collapse, the places where $\delta$ is  smaller than $\delta_R ^c$ contain less than  one Jeans mass. Eventually, these regions can either 
form a mass bigger than $M_R^c$ or form no structure at all.
Therefore, in the present context, what we are interested in is
to find out the total mass, $M_{\rm tot}(R)$, 
which  is going to form structures of mass $M_R^c$ or less.
As just mentioned, this corresponds to the places with 
a density fluctuation $\delta>\delta_R^c$.

\subsubsection{Analytical expression}
The  mass contained within structures of mass $M< M_R^c$
 is equal to the mass of the gas which, smoothed at scale $R$, has a logarithmic
density larger than $\delta_R^c$. We have
\begin{eqnarray}
M_{\rm tot}(R)  =
 L_i^3 \int ^{\infty} _ {\delta_R^c} \bar{\rho} \exp(\delta)   {\cal P}_R(\delta)  d\delta.
\label{gauche}
\end{eqnarray}
 In a way similar to the approach followed in  \S3, the mass $M_{\rm tot}(R)$ can also be 
estimated  by counting directly the self-gravitating clouds of mass {\it smaller} than $M_R^c$. 
The number of such structures   is $ {\cal N}(M') P(R,M')  dM'$,
where ${\cal N}(M') dM'$ is the density of  structures of  mass
between $M'$  and $M'+dM'$ and
$P(R,M')$ is the probability to find a gravitationally unstable cloud 
of  mass $M'$ embedded inside a cloud of gas which at scale $R$ has a logarithmic density
larger than $\delta_R^c$. 
Therefore
\begin{eqnarray}
M_{\rm tot}(R)  = L_i^3\,\int _0 ^ {M_R^c} M' \, {\cal N} (M')\,   P(R,M')\, dM'. 
\label{droit}
\end{eqnarray}
Note that the integration is from 0 to $M_R^c$ because if a self-gravitating
cloud of mass $M$ contains a self-gravitating cloud of  mass
$M'$, then, as explained previously, we assume that an object (or few objects) of mass $M'$ will form
instead of an object of mass $M$.

Estimating the probability $P(R,M')$ is not straightforward. It seems possible 
to formulate it in a way similar to Jedamzik's (1995) formulation of $P(M,M')$ in the cosmological case 
(his equations 8a and 8b). However, the complexity of the corresponding expression would lead 
to an equally complex result from which it would be difficult to extract the basic physical principles. 
On the other hand, Jedamzik finds significant deviation from the case
$P(M,M')=1/2$ only for structures 6 to 8 orders of magnitude smaller than $M_*$
\footnote{Interestingly enough, the discrepancy seems to decrease when 
 his $n$ decreases and in particular when it becomes negative, which is the case for turbulent fluctuations.}. In the present case of interest, 
the IMF typically entails masses at most 3 orders of magnitude smaller than the mean Jeans mass.  Therefore, it seems 
reasonable, in the present first step calculations, to follow the most simple approach. In the following, we thus assume that 
$P(R,M')=1$. This means that we make the assumption that any self-gravitating cloud of mass smaller than 
$M_R^c$ is embedded into a cloud which, smoothed at scale $R$, is Jeans unstable. In other words, we assume 
that the Jeans unstable clouds are not isolated but  are embedded into bigger Jeans unstable clouds (containing more 
than one Jeans mass). This assumption is further justified in Appendix \ref{justif_P=1}. 
As will be shown below by our results, this seems to be a very reasonable assumption.

Equating eqs.~(\ref{gauche}) and~(\ref{droit}) and deriving the expression with respect to $R$, one gets:
\begin{eqnarray}
\label{n_general}
 {\cal N} (M_R^c)  &=& %\\ \nonumber 
 { \bar{\rho} \over M_R^c} 
{dR \over dM_R^c} \,
\left( -{d \delta_R ^c \over dR} \exp(\delta_R^c) {\cal P}_R( \delta_R^c) + \int _  {\delta_R^c}^\infty \exp(\delta) {d {\cal P}_R \over dR} d\delta \right)
\end{eqnarray}
as the general expression for the number density distribution of star-forming collapsing regions under
the aforedefined selection criterion for collapse.

As seen, two terms appear in eq.(\ref{n_general}). Since $\delta_R^c$ is a decreasing function of  $R$, 
the first term is positive. The second term is identical to the one which appears in eq.(\ref{nclump}). 
As  shown in appendix~\ref{second_term}, this term becomes significant only when $R \simeq L_i$,  
{\it i.e.} when the size of the structures
becomes comparable to the size of the system itself. Clearly, in this limit the precise dependence of $\sigma$ on $R$ 
becomes crucial and the statistical approach becomes rather questionable. From an even more fundamental 
point of view, this  raises the question 
of the exact definition of the isolated fragmenting systems that are considered in this work,  
and how they are connected to the surrounding medium. 
We thus ignore this second term in the following, except in \S~\ref{support} where it cannot be avoided.
This implies that the spatial scale $R$ must be small compared to $L_i$ or equivalenty that the mass of the 
structures must be small compared to the mass of the system itself. 

\subsubsection{The normalization problem}
\label{norma}
In order to check whether the expression we get for the mass spectrum of collapsing structures is correctly normalized, we consider a situation where the injection scale
$L_i \rightarrow \infty$. In this limit the  system is infinite. Therefore, all its mass, $\bar{\rho} L_i^3 $,
 will eventually collapse, because even a low density piece of fluid is contained within a Jeans mass.
Then, we must have:
\begin{eqnarray}
\bar{\rho} L_i^3 = L_i^3 \int _0 ^\infty {\cal N} (M) M dM\,,
\end{eqnarray}
which, with eq.(\ref{gauche}), leads to
\begin{eqnarray}
\bar{\rho} L_i^3 = \bar{\rho} L_i^3 \int _{-\infty} ^\infty  \exp(\delta) {\cal P}_R(\delta)  d\delta,
\end{eqnarray}
It is easily shown that the integral is equal to 1 since it represents all the mass within the 
system. 
Thus, the equality is satisfied and our expression is adequately normalized.

\subsection{Mass spectrum with purely thermal support}
In case of pure thermal support, the mass spectrum of gravitationally bound objects derives from  eqs.~(\ref{form_n}),~(\ref{form_A}) and~(\ref{crit_y}), and eq.(\ref{n_general}) becomes, ignoring its second term:

\begin{eqnarray}
 {\cal N} (M_R^c)  &\simeq&  
  {2 \bar{\rho} M_J^0 \over (M_R^c)^3} {1 \over \sqrt{2 \pi} \sigma }  
 \exp \left[ - {\left[2 \ln \left( M_R^c / M_J^0 \right)\right]^2  \over  2 \sigma^2}  -{\sigma^2\over 8}   \right]
\label{therm_dens}
\end{eqnarray}
which  can also  be rewritten

\begin{eqnarray}
 {\cal N} (\widetilde{M})  \simeq
  {2 \bar{\rho} \over (M_J^0)^{2}}\, \widetilde{M}^{-3 - {2   \ln(\widetilde{M}) \over  \sigma^2} } \times {\exp( -{\sigma^2/ 8}  ) \over \sqrt{2 \pi}\, \sigma }  
\label{grav_therm2}
\end{eqnarray}
where $\widetilde{M} = M / M_J^0$.

We clearly see from eq.(\ref{grav_therm2}) that the mass spectrum involves two contributions, namely
a power-law with an index $-3$ and a lognormal contribution given by the term $- 2  \ln(\widetilde{M}) / \sigma^2 $.
The first contribution is dominant when $  M_\sigma^- \ll \widetilde{M} \ll M_\sigma^+$, where
\begin{eqnarray}
 \widetilde{M}_{\sigma}^\pm = \exp( \pm {3\over 2}\, \sigma^2 )\, ,
\label{msigma}
\end{eqnarray}
 whereas the latter one 
  eventually becomes dominant both at very large  ($M\gg M_\sigma^+$)  and at very small ($ M \ll M_\sigma^-$)
  masses, where it 
 produces an exponential cut-off. The cut-off at small masses had been  previously identified by Padoan et al. (1997).
 It arises  from both the lognormal nature of the density distribution and the threshold
condition for collapse (eqs.\ref{crit_y}, \ref{crit_yt}, \ref{crit_ytot}),
 as demonstrated in the present paper. This provides a rigorous foundation for this peculiar
form of the power-law exponent of the CMF/IMF, sometimes invoked empirically in the literature \cite{MS79}.

This clearly demonstrates
that the stellar CMF/IMF results from two contributions, a power-law which dominates in the aforementioned mass range,
 and a lognormal form, which becomes important at very small and very large masses, as characterized
  by the transition mass
 $\widetilde{M}_{\sigma}^\pm$ (eq.\ref{msigma}).
Equation (\ref{grav_therm2}) also shows that the mass spectrum of bound objects issued from a
purely thermal collapse has a much steeper distribution at large masses, $ {\cal N} \propto M^{-3}$, than the one given by the
Salpeter CMF/IMF, ${\cal N}(M) \propto M^{-2.35}$. It also highlights the importance of the characteristic
scale of the system on the mass spectrum, through the scale-dependence of the variance $\sigma$  (see \S2). 
 Finally it also clearly shows the importance of the Mach number on the mass spectrum of collapsing prestellar cores. Indeed, with eqs.(\ref{sigma_val}) and~(\ref{msigma}), we get:
\begin{eqnarray}
 \widetilde{M}_{\sigma}^\pm = (1 + b {\cal M}^2)^{\pm {3\over 2}} , 
\label{msigma1}
\end{eqnarray}
Small-scale motions, i.e. small values of $\sigma$, will hardly produce any object far away from the mean Jeans 
mass.

\subsection{Mass spectrum with purely turbulent support}

In case the clumps are supported dominantly by turbulent motions, eqs.~(\ref{form_n}),~(\ref{form_A}), 
~(\ref{crit_yt}) and~(\ref{n_general}) yield

\begin{eqnarray}
 {\cal N} (\widetilde{M})  &\simeq&  
{2 \bar{\rho} \over (M_J^0)^{2}}\,  { (1- \eta)  \over (2 \eta +1)}\, {\cal M}_*^{3 / (2 \eta+1)} \, \widetilde{M} ^{- 3\alpha_1 }   
\times \exp \left(  - {\left( \ln ({\cal M}_* ^{\alpha_3} 
\widetilde{M} ^{2\alpha_2} )  \right)^2 \over 2 \sigma^2}   \right)  
\times {\exp( -{\sigma^2/ 8} ) \over \sqrt{2 \pi}\, \sigma } \nonumber \\
&=& {2 \bar{\rho} \over (M_J^0)^{2}}\,  { (1- \eta)  \over (2 \eta +1)}\, {\cal M}_*^{6 / (\eta-1)} \,
\, \widetilde{M^\prime} ^{- 3\alpha_1 - {2 (\alpha_2)^2 \over  \sigma^2}    \ln(\widetilde{M^\prime}) }
\times {\exp( -{\sigma^2/ 8} ) \over \sqrt{2 \pi}\, \sigma },
\label{grav_turb2}
\end{eqnarray}

\noindent where $\widetilde{M} = M / M_J^0$, $\widetilde{M^\prime}={\cal M}_*^{3 / (\eta-1)}\widetilde{M}$,
%\begin{eqnarray}
$\alpha_1 =  (1 + \eta) / (2 \eta +1) $,
%\end{eqnarray}
%\begin{eqnarray}
$\alpha_2 = ( \eta-1) / (2 \eta +1 )$,
%\end{eqnarray}
%\begin{eqnarray}
$\alpha_3 = 6 / (2 \eta +1 )$,
%\end{eqnarray}
and
\begin{eqnarray}
%{\cal M}^2_* = { V_0 ^2  \over 3 C_s^2} 
%\left({\lambda_J^0 \over  L_i}\right) ^{2 \eta}.
{\cal M}_* = { 1  \over \sqrt{3} } { V_0  \over C_s}\left({\lambda_J^0 \over   1 {\rm pc} }\right) ^{ \eta}
\approx (0.8-1.0) \,\left({\lambda_J^0\over 0.1\,{\rm pc}}\right)^{\eta}\,\left({C_s\over 0.2\kms}\right)^{-1},
\label{mach_eff}
\end{eqnarray}
%and where as in the previous case, we have neglected the second term.
where the ratio $(V_0/1\, {\rm pc})^\eta$ is given by the aforementioned Larson relation (eq.\ref{larson}).
Roughly speaking, the effective Mach number ${\cal M}_*$ measures the relative importance of
turbulent versus thermal support contributions  at the Jeans length scale.
Note that, according to the discussion of \S4.4, this effective Mach number ${\cal M}_*$ can be renormalized to take into account the presence of a magnetic field. This requires, however, a proper knowledge of the exact dependence of
both the uniform and fluctuating components of the magnetic field (or Alfv\'en velocity) upon the thermal and non-thermal contributions of the velocity dispersion, respectively.

From comparison between eqs.(\ref{grav_therm2}) and (\ref{grav_turb2}), we first note that the introduction
of the turbulent contribution into the effective sound speed modifies the exponent of the power-law
term, through the Larson exponent. Interestingly enough, the aforementioned favored value for
supersonic turbulence, $\eta=0.4$-0.45 (Kritsuk et al. 2007) yields {\it exactly} the Salpeter coefficient, $d N / d M \propto M^{-(1+x)}$. Indeed,
$1+x=3\alpha_1=2.33$-2.35, bracketed by the Burgers, $3\alpha_1=2.25$,
and Kolmogorov,  $3\alpha_1=2.4$, values. Expressed as a function of the (3D) index of turbulence,  $n$,
with the help of relation (\ref{eta}), the
power law exponent of the mass spectrum as obtained in our calculations reads:
\begin{eqnarray}
x = { n + 1 \over 2 n - 4 }.
\label{turb_salpeter}
\end{eqnarray}

As for the thermal case, the precise value of the turnover mass,
around which the CMF/IMF evolves from the power-law to the lognormal form, depends on the value of $\sigma$
and therefore on  the Mach number ${\cal M}$. Larger Mach values will produce larger numbers of small-scale collapsing
clumps.

 The reason why  the  mass spectrum is stiffer when thermal support only is considered
than when turbulent support is taken into account is simply because the turbulent 
support increases with the scale. Thus, a lot of intermediate to relatively large
 mass structures (of the order of or larger
than the usual Jeans mass) which are unstable under 
purely thermal criteria are stabilized by turbulence. This support thus prevents fragmentation of these
structures into several smaller structures, leading naturally to a shallower and broader mass spectrum
in the high mass ($M>M_J^0$) domain. 
%Therefore, when turbulence support is  taken into account, the mass spectrum is  shallower. 

\subsection{General case}

In the general case, where both thermal and non-thermal supports contribute, the mass spectrum now
reads, from eqs.(\ref{form_n}),~(\ref{form_A}),~(\ref{crit_ytot}), 
~(\ref{crit_Mtot}) and eq.(\ref{n_general}) where, as explained previously, the second term has been dropped:
\begin{eqnarray}
{\cal N} (\widetilde{M} ) &\simeq& 2\, {\cal N}_0 \, 
{ 1 \over \widetilde{R}^3}  {1 \over 1 + (2 \eta + 1) {\cal M}^2_* \widetilde{R}^{2 \eta} } 
 \left(
{ 1 + (1 - \eta){\cal M}^2_* \widetilde{R}^{2 \eta} \over
 \, (1 + {\cal M}^2_* \widetilde{R}^{2 \eta})^{3/2}} 
\right) \nonumber
\\ &\times& 
\exp \left( - { \left[ \ln \left( { \widetilde{M}  /
\widetilde{R}^3 }  \right) \right]^2 \over  2 \sigma^2 } \right)
\times {\exp( -\sigma^2/8 ) \over \sqrt{2 \pi} \sigma } 
\label{big_dens}
\end{eqnarray}
where $\widetilde{R}= R / \lambda_J^0$,
$\widetilde{M} =  M / M_J^0 = \widetilde{R}\,
(1+ {\cal M}^2_* \widetilde{R}^{2 \eta})$, $\delta_R^c = \ln \bigg\{ (1 + {\cal M}^2_* \widetilde{R}^{2 \eta}) / \widetilde{R}^2  \bigg\} $ and ${\cal N}_0=  \bar{\rho} / (M_J^0)^{2}$.

This expression can also be rewritten:
\begin{eqnarray}
{\cal N} (\widetilde{M} ) &=& 2\, {\cal N}_0 \, { 1 \over \widetilde{R}^6} \,
{ 1 + (1 - \eta){\cal M}^2_* \widetilde{R}^{2 \eta} \over
[1 + (2 \eta + 1) {\cal M}^2_* \widetilde{R}^{2 \eta}] }
\times    \left( {\widetilde{M} \over \widetilde{R}^3}  \right) ^{-{3 \over  2} -   {1 \over 2 \sigma^2} \ln (\widetilde{M} / \widetilde{R}^3) }
\times {\exp( -\sigma^2/8 ) \over \sqrt{2 \pi} \sigma },
\label{grav_tot2}
\end{eqnarray}

We see that the transition between the thermal and the turbulent dominated regimes, 
occurs when the radius $\widetilde{R} \simeq {\cal M}_*^{-1/\eta}$ and thus for masses around
\begin{eqnarray}
\widetilde{M}^{*} \simeq 2 ({\cal M}_*) ^{-1/\eta}. 
\label{m_et}
\end{eqnarray}
For ${\cal M}_* ^2 \simeq 2$, we get $\widetilde{M}^{*}\simeq 0.8-1$.     
At masses larger than $\widetilde{M}^{*}$, we recover the power-law behaviour characteristic 
of the turbulent collapse, with the
proper Salpeter value. This is easily verified for the case $\eta=0.5$, 
which yields ${\cal N} (\widetilde{M} )_{\widetilde{M}\gg 1} \propto  \widetilde{M}  ^{-9 /  4 }$.
The small-mass limit (i.e. $\widetilde{M} \ll \widetilde{M}^*$), on the other hand, resembles the 
one of the purely thermal case. 
Therefore, at least for values of ${\cal M}_*$ of the order of unity, we expect
the mass spectrum of collapsing structures in the general case to be bracketed by the turbulent and
thermal behaviours at large and small scales, respectively.

\section{Results}
In this section, we study  eq.(\ref{grav_tot2}) and its dependence upon the parameters, 
$\cal M$ (which enters the expression of $\sigma$) and $ \cal M_*$. Recalling that  the Mach number 
$\cal M$ is the ratio of  velocity dispersion over sound speed at the scale of  the {\it whole cloud}, whereas  
$\cal M_*$ is the ratio of  velocity dispersion over  sound speed at the scale of the {\it Jeans length}, 
we see that both parameters depend on the velocity dispersion. The first one increases with 
the size of the cloud while the second one increases with the size of the Jeans length. 
In order to investigate their respective influence  on the CMF/IMF, we will first vary one of these two
parameters while keeping 
the other one constant. Physically speaking, this corresponds to either considering clouds of fixed 
density but of various sizes (${\cal M}$ varies but not ${\cal M}_*$), or changing 
the density but not the size (${\cal M}_*$ varies but not ${\cal M}$). 
For a sake of simplicity, we assume in these 2 cases that $\sigma=\sigma_0$.

Finally, in order to investigate the 
influence of turbulence on star formation, we will vary simultaneously 
${\cal M}$ and ${\cal M}_*$ while keeping their ratio constant. This latter case corresponds to a cloud 
of fixed size and density but of varying velocity dispersion. In this case, the dependence of $\sigma(R)$
on $R$ is properly taken into account.

The results are expressed in terms of $M_J^0$ as defined by eq.(\ref{jeans}). 
Typically, for a cloud of density $\bar{n} \simeq 10^3$ cm$^{-3}$,  $M_J^0$ is about 3 $M_\odot$.
For clouds  of size $\simeq 1-2$ pc following the Larson laws, 
 typical Mach number values are ${\cal M}\sim 6$.   Typical values of ${\cal M}_*^2$ are around 1-2. 
These numbers must be considered as indicative, given the large uncertainties that remain on the 
exact conditions for star formation. 
In the following, the value of $\eta$ is taken to be equal to 0.4.

\subsection{Thermal and turbulent support }
\begin{figure}[p]
\center{\includegraphics[angle=0,width=6in]{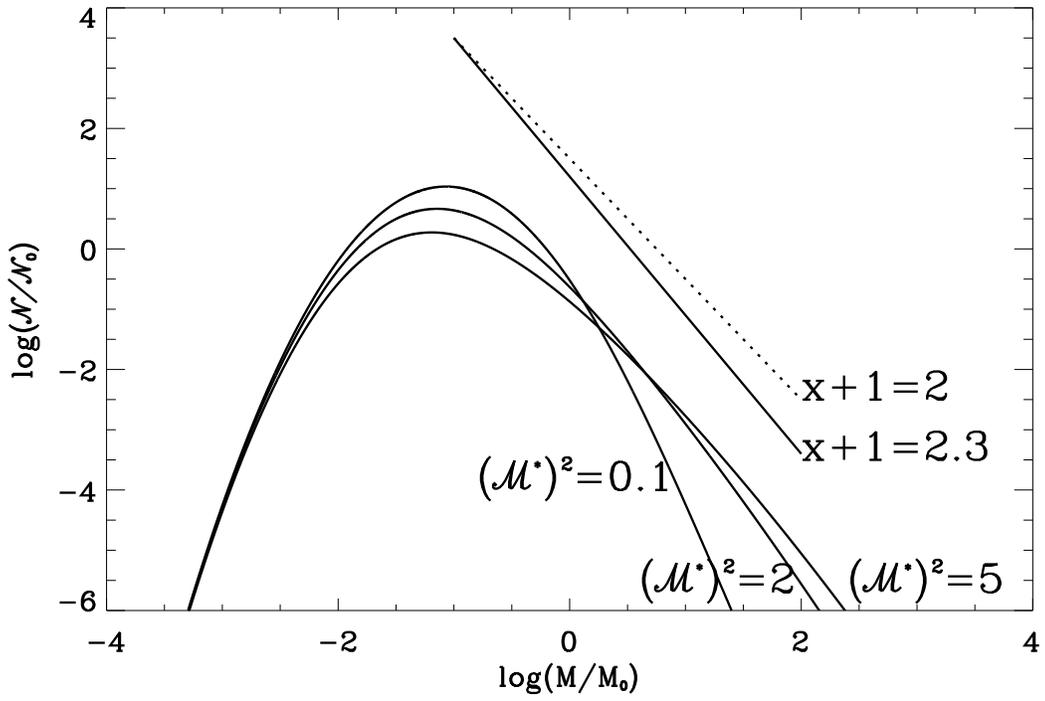}} 
\caption{Mass spectrum for ${\cal M}=6$ and various values of ${\cal M}_*^2$.
For reference, power-law distributions $dN/dM \propto M^{-(x+1)}$ are also plotted for $x+1=2$ and 
$x+1=2.3$, about the Salpeter value.}
\label{fig_mach_et}
\end{figure}

Figure~\ref{fig_mach_et} portrays the mass spectrum obtained in the general case (eq.(\ref{grav_tot2})),
for different values of ${\cal M}_*$, namely ${\cal M}_*^2=0.1$, 2 and 5, at a fixed Mach number 
${\cal M}=6$. 

For the first value of ${\cal M}_*$, the influence of the turbulent support is negligible ($\widetilde{M}^*\simeq 30$) and the collapse is almost purely thermal. 
We see the steep cut-off at large masses due to the $\alpha=-3$
exponent in the power-law, as well as the sharp exponential cut-off at small masses. As mentioned
above, this
latter term  derives from the general formalism and from the critical density selection criterion, and thus
is {\it not} specifically due to turbulence, but to the lognormal distribution characteristic of
the density field.

As expected, for ${\cal M}_*^2=2$ the behaviour at small masses is 
almost identical to the purely thermal case. The CMF/IMF peaks at almost the 
same position, although less intermediate  mass stars form because of the turbulent support, which prevents the collapse of large structures into smaller ones. 
At large masses, the CMF/IMF is less stiff than for the pure thermal case and 
has an index close to the Salpeter value,as mentioned earlier. 
In the case ${\cal M}_*^2=5$, the number of intermediate mass stars
is even smaller, but the peak of the CMF/IMF remains weakly affected (as expected from eqn.(\ref{m_et})).

\subsection{Influence of the Mach number ${\cal M}$}

\begin{figure}[p]
\center{\includegraphics[angle=0,width=6in]{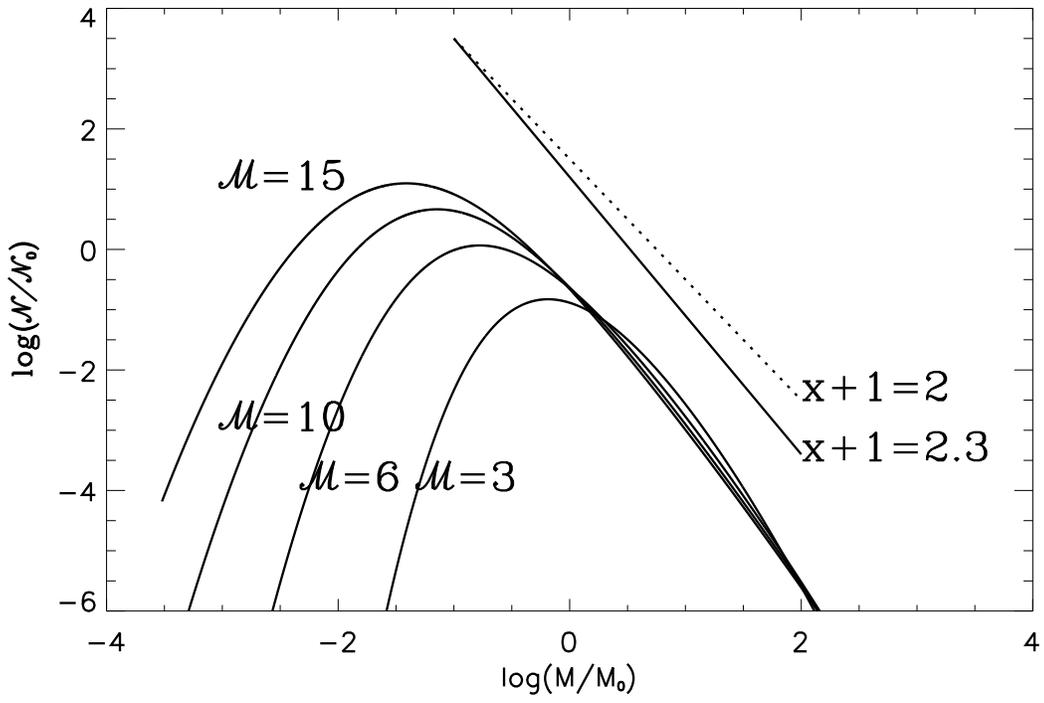}} 
\caption{Mass spectrum for ${\cal M}_*^2=2$ and various values of $\cal M$.}
\label{fig_mach}
\end{figure}

 Figure~\ref{fig_mach} shows the CMF/IMF for ${\cal M}_*^2=2$  and various values 
of the Mach number $\cal M$. We stress that, since ${\cal M}_*$ is maintained constant, 
increasing the Mach number is not equivalent to increasing the value of $V_0$, but rather the
size of the cloud. 
 
 Clearly, supersonic turbulence strongly enhances the collapse
of small-scale structures while the number of large-scale structures
does not change significantly (for these values of ${\cal M}$).
This stems from the fact that the transition mass, $M_\sigma^-$, defined by eq.(\ref{msigma}), is modified
by the variance $\sigma$, so that the exponential
cut-off occurs at much smaller masses, for a given hydrodynamic Mach number ${\cal M}$, than for
the thermal case. 

This shows that, if large-scale motions in molecular clouds are strongly supersonic,
they will produce a large population of brown dwarfs. It is the fact that
small-scale structures become subsonic (eq.\ref{larson}), and thus of dominantly thermal nature, that they
hardly form gravitationally bound objects far away from the mean Jeans mass, and thus that
brown dwarfs, although issued from the same general formation mechanism, are not formed as efficiently as (low-mass) stars around the mean Jeans mass.  
In the high-mass regime, as mentioned
previously, turbulence yields a shallower power-law tail than for thermal collapse. Mach number values
$\cal M\sim $ 6 yield the correct Salpeter index already for $M\ga M_J^0$,  for a power spectrum index $n\simeq 3.8$ ($\eta \simeq 0.4$), as inferred from recent high resolution simulations of supersonic turbulence (Kritsuk et al. 2007).
As expected, the general case indeed lies between the turbulence
dominated collapse behaviour at large masses and the thermal behaviour at small scales. Not surprisingly, the characteristic mass
of the spectrum is fixed by the characteristic mass for collapse, affected to some
extent by the Mach number (which enters the variance), illustrating the role played
by gravity in this collapse gravo-hydrodynamical picture.

\subsection{Support versus turbulent triggering}
\label{support}
\begin{figure}[p]
\center{\includegraphics[angle=0,width=6in]{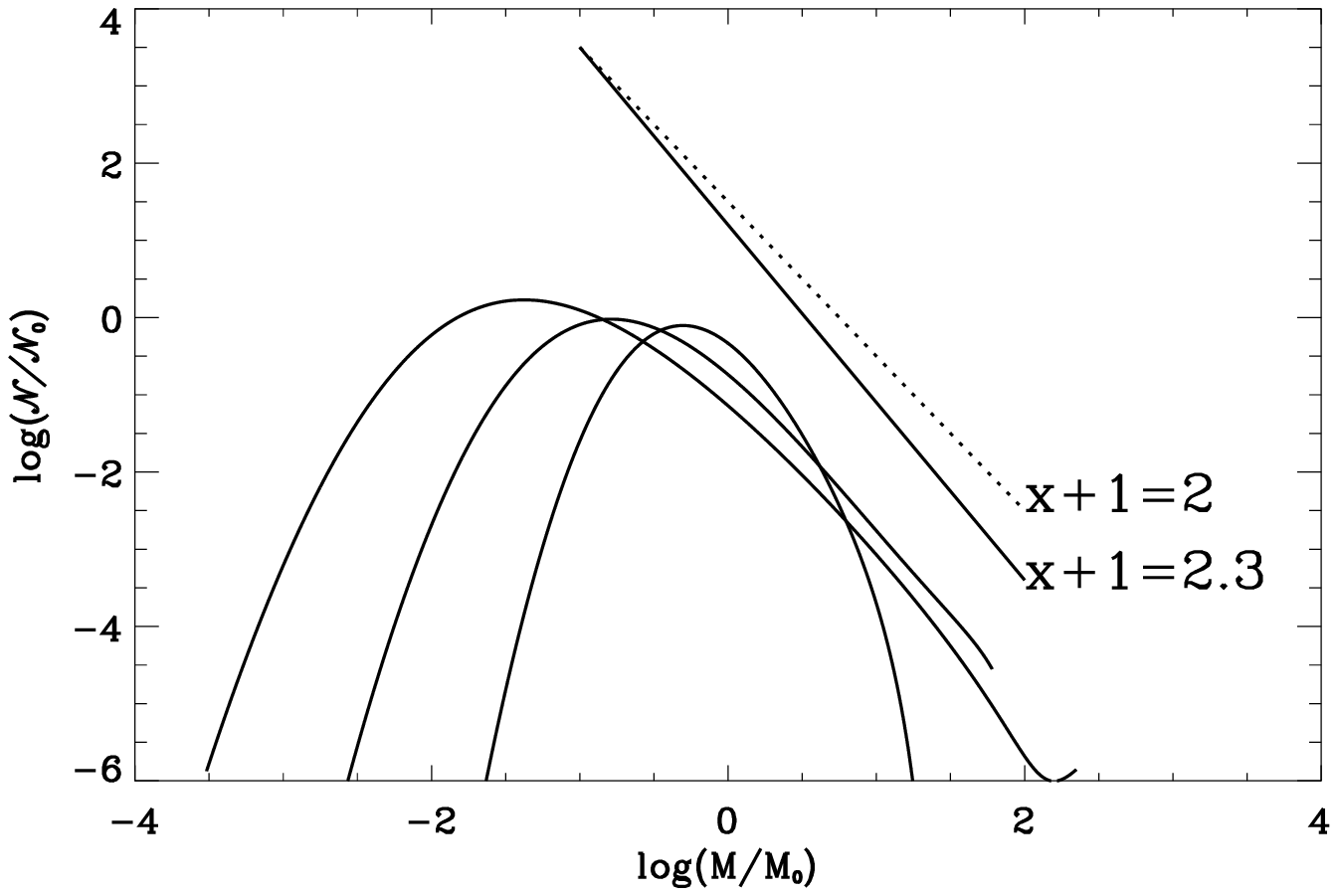}} 
\caption{Mass spectrum for various values of $\cal M$ and ${\cal M}_*$ for a constant ratio ${\cal M}/{\cal M}_*=4.24$. 
Values of $\cal M$ correspond to 3, 6 and 12 from the narrower to the broader distributions,
respectively.}
\label{fig_mach_synchro}
\end{figure}

A key, yet unsettled issue in the modern picture of star formation is to clearly understand the overall
effect of turbulence on this process.
On one hand, turbulence creates density enhancements (obviously related to $\cal M$,
 as given by eq.\ref{msigma1}) which 
tend to favor gravitational instability. On the other hand, the turbulent support tends to 
stabilize the gas (the turbulent support is due to ${\cal M}_*$ in our theory).
Figures~\ref{fig_mach_et} and~\ref{fig_mach} show how the CMF/IMF changes with 
$ {\cal M} $ and $ {\cal M} _*$. However, if one increases the level of turbulence in 
a cloud of fixed size and mass, both parameters should change, while their ratio should remain constant. 
Figure~\ref{fig_mach_synchro} displays the mass spectrum obtained for three values of 
$\cal M$, namely ${\cal M}=3$, 6 and 12, while keeping ${\cal M} / {\cal M}_*={\sqrt 3}(L_i/\lambda_J^0)^\eta$ constant and equal to $6/\sqrt{2}$ 
(i.e. $\lambda_J^0/L_i\simeq 0.1$),
leading to ${\cal M}_*^2=0.5$, 2 and 8, respectively.  
As expected, we obtain a combination of the behaviours observed in Fig.~\ref{fig_mach_et}
and~\ref{fig_mach}: the smaller the values of $\cal M$ and ${\cal M}_*$ the narrower the mass
range of star formation around the characteristic mass $M_J^0$. Conversely, large Mach numbers promote the formation of
a larger number of low-mass and high mass stars but produce less stars around $M_J^0$. 

\begin{figure}[p]
\center{\includegraphics[angle=0,width=6in]{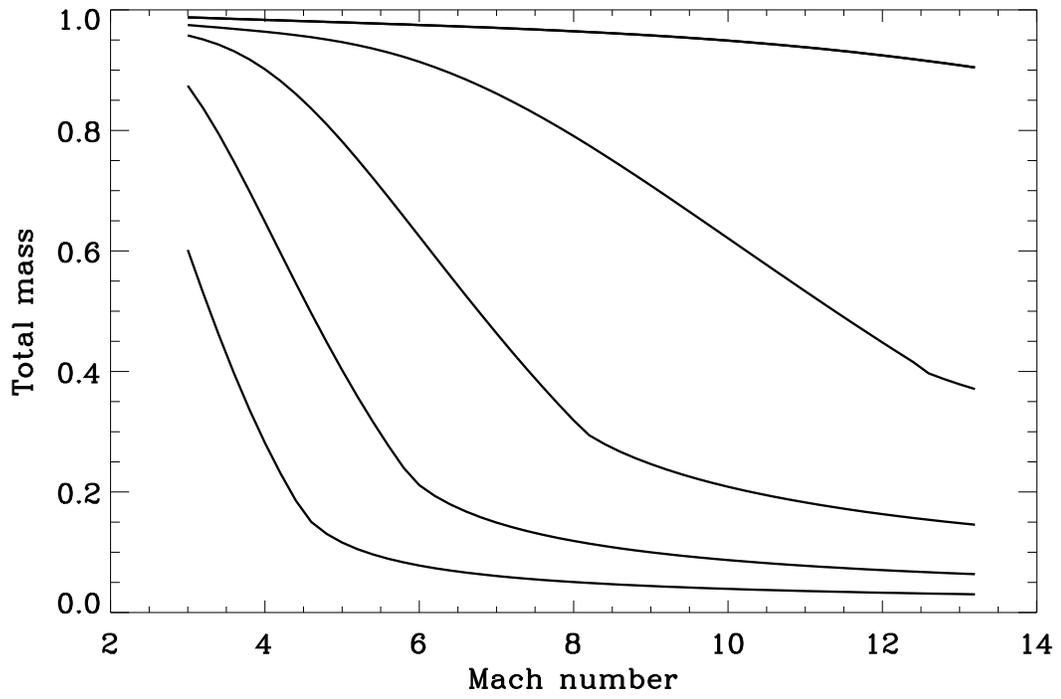}} 
\caption{Ratio of the  mass within stars over the initial mass as a function 
of the Mach number ${\cal M}$ for various values of ${\cal M} / {\cal M}_*$ (see text) (decreasing from top to bottom). }
\label{fig_mass}
\end{figure}

Another important issue is to determine whether turbulence is globally 
promoting or quenching star formation. 
 Numerically (e.g. MacLow \& Klessen 2004), it has been found that 
the global effect of turbulence on star formation is negative, although the turbulent support 
is much more effective if the driving is imposed at small scales. 
To verify this conclusion, we have
computed the total mass included in the gravitationally bound prestellar cores, $M_{tot}=\int  {\cal N} 
(M) M dM$, for various 
values of $\cal M$ and various values of ${\cal M} / {\cal M}_*$, namely 
 $6/\sqrt{2}$, $6/\sqrt{3}$, $6/\sqrt{4}$ and  $6/\sqrt{5}$, which correspond to $\lambda_J^0/L_i
\simeq$ 0.1, 0.18, 0.25 and 0.33, respectively.
For these calculations, we have included the previously neglected second term in eq.(\ref{n_general}) 
and thus use the complete relation for ${\cal N}(M)$. 
As explained in \S\ref{norma},  the integral $\int_0^\infty M {\cal N} (M) dM$ is constant. 
In reality, however, the system has a finite mass and size, so that the integral must be truncated
when $R \simeq L_i$ or, equivalently, when $M \simeq \bar{\rho} L_i^3$. Thus, the value of the integral 
depends on the injection scale.  As discussed previously, integrating until $R = L_i$ is 
questionable, and we thus stop  the integration at $R = 2 L_i/3 = L_{\rm cut}$. We have verified that using different values of $L_{\rm cut}$ yields the same trends and 
qualitatively similar results. Quantitatively speaking, however, the results obviously depend on the value 
of $R$ at which the integration is stopped.

Figure~\ref{fig_mass} portrays the results of this global star formation efficiency as a function of the Mach
number ${\cal M}$. The top curve corresponds to the largest value
of  ${\cal M} / {\cal M}_*$, i.e. the smallest value of  $\lambda_J^0/L_i$, whereas the bottom curve corresponds to the opposite. 
We see that, for a fixed value of ${\cal M} / {\cal M}_*$, the higher the Mach number
the smaller the star formation efficiency. In the same vein, for a given Mach number, 
the higher  ${\cal M}_*$ the less efficient the star formation. 
These behaviours are in agreement with and provide a theoretical foundation to
what has been inferred from numerical simulations. We stress that, quantitatively, these results depend on 
the choice of $L_{\rm cut}$. They demonstrate, however, that turbulence largely decreases the efficiency of star formation.

%We note that the departure of $M_{tot}$ from 1 is due to the CMF/IMF being 
%truncated at some mass value $\simeq L_{\rm cut}^3 \bar{\rho}$. When the total mass of star forming cores 
%is well below 1, this means that 
%the truncation occurs at masses not very different from $\widetilde{M}=1$.

\section{Discussion}

\subsection{Comparison with observations}

\setlength{\unitlength}{1cm}
\begin{figure}
\begin{picture}(0,15)
\put(0,0){\includegraphics[angle=0,width=9cm]{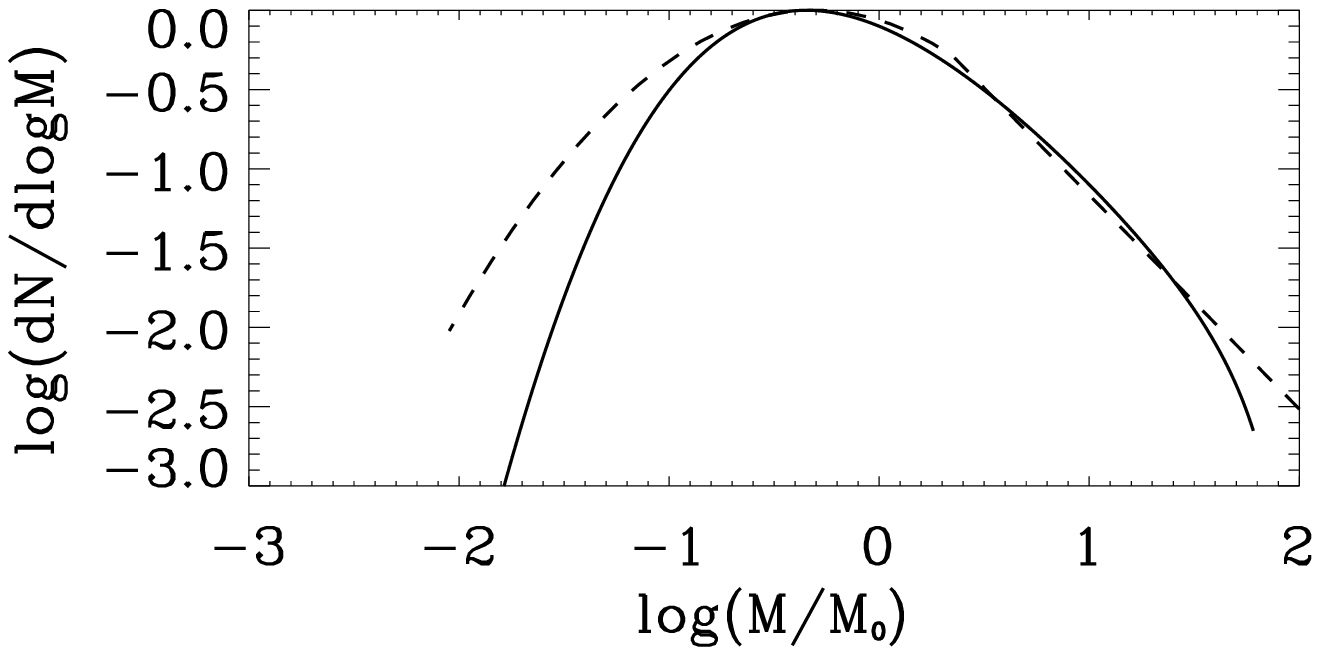}} 
\put(0,5){\includegraphics[angle=0,width=9cm]{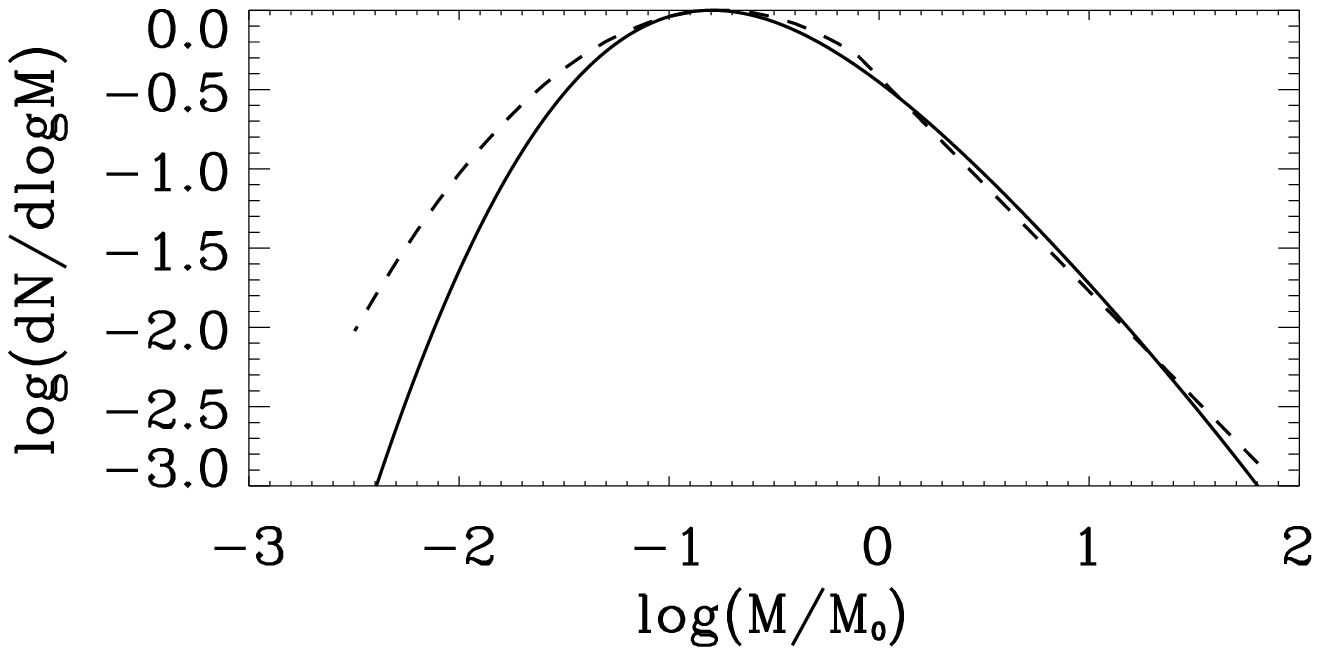}} 
\put(0,10){\includegraphics[angle=0,width=9cm]{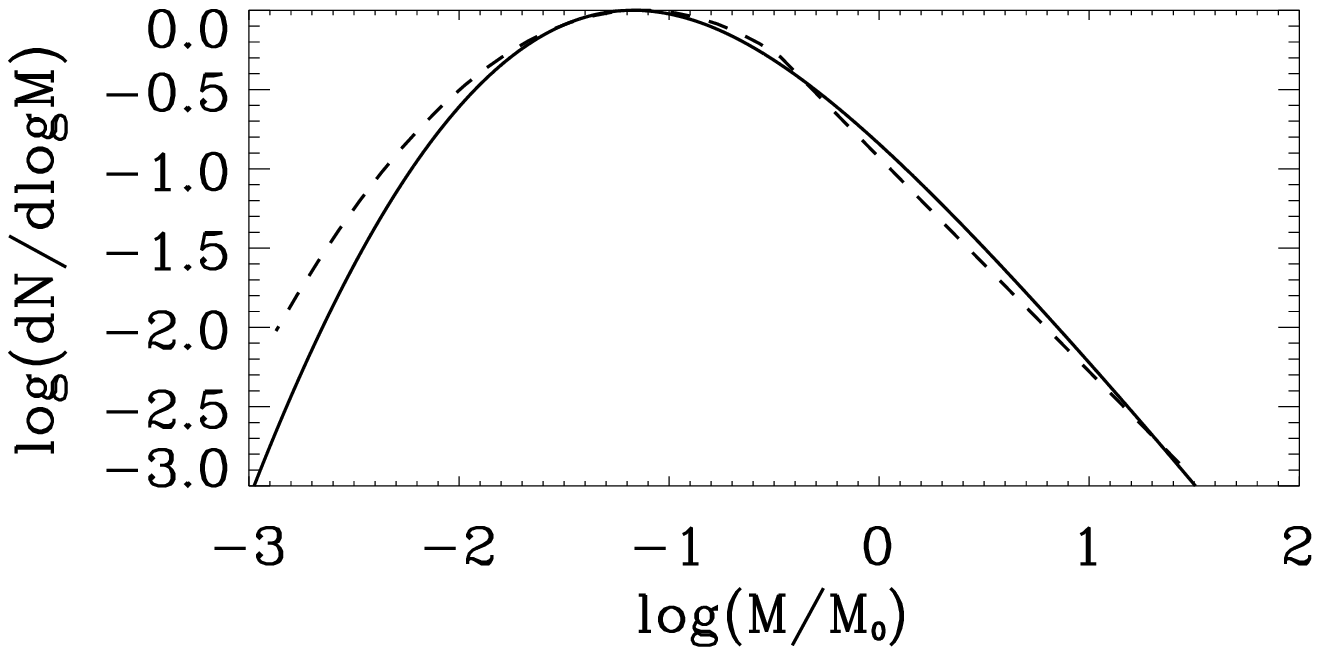}} 
\end{picture}
\caption{Comparison between the theoretical IMF/CMF, $dN/d\log M$, (solid line) obtained with  
${\cal M}=6 $ (lower panel), ${\cal M}=12 $ (middle panel), ${\cal M}=25 $ (upper panel)
  and  ${\cal M}_*^2=2$ and the stellar/brown dwarf {\it system}
initial mass  function (dotted line) of Chabrier (2003a). The peak of the this latter IMF has been adjusted
arbitrarily to the one of each theoretical mass function.}
\label{comparaison}
\end{figure}

\subsubsection{The shape of the CMF/IMF}

Figure~\ref{comparaison} compares our analytical CMF/IMF (eq.\ref{grav_tot2}) for 3 
values of the Mach number, namely ${\mathcal M}=6$,
12 and 25, with the IMF representative of the Galactic field and young 
clusters (Chabrier 2003a). This latter reflects the so-called {\it system}
IMF since, in general, present limitations on angular resolution do not 
allow to resolve these systems into individual objects. On the other 
hand, our calculations are representative of the early stages of the 
collapse and do 
not consider possible subsequent
subfragmentation of the prestellar cores into multiple objects. The same 
is true for most of present numerical
simulations of star formation. In order to facilitate the comparison between
the theoretical and the observationally derived distributions, this 
latter has been shifted in order to match the position of the peak of 
the analytical distribution. We will come back to this point below. The 
general agreement between the
two types of distributions is striking. In particular, the transition 
between the power-law tail and the lognormal form is very well 
reproduced. This clearly assesses the validity of such a
composite functional form for the stellar IMF, whose physical foundation has been 
demonstrated in the present paper.
At the low-mass end, the theoretical and
observationally-derived mass spectra start to differ noticeably below 
some mass value which depends on the Mach number,
as analyzed in the previous section.
A value ${\mathcal M}=6$ leads to a deficit of both very massive stars 
and brown dwarfs compared with the observed distribution, because of the 
too restricted mass range between the high-mass and the low-mass exponential
cut-offs. A value ${\mathcal M}=12$ improves the
situation and yields a fairly reasonable agreement with the 
observational distribution, although still underestimating the number of 
very small ($\la 10^{-2}\,\msol$) brown dwarfs.
A value ${\mathcal M}=25$ yields a good agreement with the 
observationally derived mass spectrum
over basically all the observationally probed mass domain. Although values
${\mathcal M} \simeq $ 25 are substantially larger than the ones usually 
associated with typical star forming regions, ${\mathcal M}\sim 6$-12, 
two important points need to be considered in these comparisons.

First of all, the statistics of brown dwarf detections below $\sim 
10^{-2}\,\msol$ are still small and determinations of the
brown dwarf densities remain hampered by both observational  
(magnitude-limited samples, spectral-type effective temperature 
conversions, etc...) and theoretical (mass-age-$L$-$T_{eff}$ 
determinations) uncertainties. The observationally-derived IMF in this 
domain thus retains a large degree of uncertainty.

Second of all, the relation between the variance of the PDF obtained by 
turbulence, which determines the
width of the CMF/IMF, and the Mach number is given by 
eq.(\ref{sigma_val}), with $b=0.25$. This relationship
is extracted from 3D simulations of {\it isothermal} turbulence (Padoan 
\& Nordlund 2002).
Simulations taking into account a detailed treatment of the thermal 
properties of the gas in the molecular cloud, however, lead to an 
equation of state softer than isothermal, $P\propto \rho^{0.7-0.8}$, at 
least up to densities of a few $10^3$ cm$^{-3}$  (Glover \& MacLow 2007). We thus 
expect a larger dispersion of the PDF and thus a larger
extension of the CMF towards small
masses with a more compressible non-isothermal gas, for the same Mach 
number. Furthermore, recent numerical
simulations of supersonic turbulence with compressible driven modes 
predict a significantly broader PDF, for the same rms Mach number, than  
simulations with rotational driven modes,
 like the ones of Padoan \& Nordlund (W. Schmidt, priv. com.).  
These results  suggest that the relation 
(\ref{sigma_val}) underestimates somehow the aforementioned value of $b$ 
and thus the variance $\sigma$ for a given ${\mathcal M}$.
Finally, self-gravity could also possibly broaden the density PDF, as supported by 
numerical simulations (see e.g. V\'azquez-Semadeni et al. 2008), although the densest 
part of the density distribution in that case departs from a pure lognormal form.

Consideration of these issues should improve the quantitative agreement
between the present theory and observations.

\subsubsection{From the CMF to the IMF: the core-to-star formation efficiency.}

As mentioned above, the theoretical and observationally-derived mass 
functions, $d{\mathcal N/}d\log M$, have been arbitrarily adjusted peak-to-peak. 
After the shape of the CMF/IMF, we now examine the agreement between the 
real positions of the peaks of the
CMFs, which define the most probable stellar mass scale. 
The comparison with the 
observationally-derived IMF on a solar-mass scale, however, depends on
the exact value of the Jeans mass which, in our approach, defines the 
characteristic mass for collapse. Given the fact that the observationally derived system IMF peaks 
at roughly 0.3$M_\odot$, this would imply that the Jeans mass should be about 
1$M_\odot$ for ${\cal M}=6$ and $\sim 3$ $M_\odot$ for ${\cal M}=25$.

On the other hand, observations of
prestellar condensations, as identified in dust continuum surveys, show 
that the CMF peaks at a mass
about three times larger than the observational IMF, even though this 
peak determination still suffers from large
uncertainties (Motte et al. 1998, Alves et al. 2007, Andr\'e et al. 2007, 2008).
Given the similarity of the shapes of the CMF and the IMF, this implies 
that the transformation of gravitationally bound prestellar condensations
into genuine stellar or brown dwarf systems involves some {\it uniform} star
formation efficiency factor, $\epsilon=M_\star/M_{core}$, observed to be 
in the range $\epsilon \sim 30$-50\%.
Clearly, the process(es) 
responsible for
the evolution of the CMF into the IMF remains ill-determined, although 
magnetically-driven outflows,
expected to produce a mass-independent star formation efficiency factor 
in the appropriate range (Matzner \& McKee 2000), offer an appealing 
explanation.
This  implies that the above estimate for the Jeans mass value adequate to match 
the observed IMF must be further increased by a factor of the order of $\sim 3$,
yielding $M_J^0\simeq $ 3 $M_\odot$ for ${\cal M}=6$ and $M_J^0\simeq$ 10 $ M_\odot$ for ${\cal M}=25$.

\subsubsection{Physical conditions}

Equation (\ref{jeans}) shows that, for star forming core conditions
$\mu=2.33$, $T=10$ K,
$\bar n=10^3$ cm$^{-3}$, the Jeans mass is about 3 $M_\odot$, whereas it is about 1$\msol$ for
$\bar n=10^4$ cm$^{-3}$.
Fluctuations around these temperature and density values, as well as 
geometrical factors,
however, will affect this determination to some degree, moving around 
the peak of the theoretical
mass spectrum on a solar-mass scale.
This would be consistent with the IMF being determined at gas densities
in this density range.
%of about $10^3$ cm$^{-3}$.
However, as discussed in \S4.4, the presence of a magnetic field
increases the critical mass for collapse, moving the peak of our CMF, 
normalized to the standard thermal Jeans mass
(eq.\ref{jeans}), to larger masses.
Estimates of the critical magnetized mass for collapse, 
as done in various seminal works and textbooks
% (e.g. Mouschovias \& Spitzer 1976, Shu et al. 1987, Lequeux 2005),
(e.g. Spitzer 1978, Lequeux 2005), 
predict values for the relevant magnetic Jeans mass, $M_{\Phi}\propto B^3/n^2$, of the order of $\sim 10\,\msol$
for $\bar n=10^3$ cm$^{-3}$ and $B=10\,\mu$G. This implies that 
the typical gas density to produce a magnetic Jeans mass compatible
with a peak of the CMF around 1 $M_\odot$, as found observationally, should be higher than $10^3$ cm$^{-3}$,
with a typical value of the magnetic field slightly larger than 10 $\mu$G. This
is consistent with the densities characteristic of star forming clumps, ${\bar n}\sim 10^4-10^5$ cm$^{-3}$ 
(Andr\'e et al. 2007, Motte et al. 2007). 
On the other hand, the nice correlation observed by Basu (2000) suggests that the 
magnetic field also depends on the velocity dispersion. According to the analysis of
\S4.4, this implies that only the {\it uniform} magnetic component should enter
our Jeans mass determination. Unfortunately, the value of this component is not determined observationally.
Clearly, more knowledge on the structure of the magnetic field 
and its correlation with density and velocity is required at this stage  in order to determine more precisely the value of the
magnetic Jeans mass and to make a precise comparison between the characteristic
mass scales, thus peak positions, of the theoretical and observationally derived IMFs.

\subsubsection{Expected dependence of the peak position in the cloud mass}
One surprising observational evidence, is that the peak of the IMF is fairly constant 
from one region to another (e.g. Chabrier 2003a, 2005).

Here, we investigate how the peak of the IMF depends on the mass of the cloud.
Since, as emphasized earlier, the peak occurs in the thermally dominated regime, we 
simply take the derivative of $ {\cal N} $ as given by eq.(\ref{grav_therm2}), with respect to 
$M$, which yields:
\begin{eqnarray}
 \widetilde{M}_{\rm peak} = \exp ( - {3 \over 4 } \sigma ^2 ) = { 1 \over (1 + b {\cal M}^2) ^{3/4}}.
\label{mpeak}
\end{eqnarray}
This indicates that, for clouds characterized by high Mach numbers, the peak of the core mass function is roughly proportional 
to ${\cal M}^{-3/2}$. Combined with eq.(\ref{larson}), this gives  $\widetilde{M}_{\rm peak} \propto L^{-3/2  \eta}$. 
Since, typically, the CO clump density varies with their size as $\rho \propto L^{-a}$,
with $a=0.7-1$ (Larson 1981, Heithausen et al. 1998), we get
\begin{eqnarray}
{M}_{\rm peak} = M_J^0 \widetilde{M}_{\rm peak}  \propto L ^{(-3/2) \eta + a / 2} 
\label{mpeak2}
\end{eqnarray}
For $\eta=0.5$, $a=1$ or for $\eta = 0.4$, $a=0.7$, we find the same value 
$(-3/2) \eta + a /2 = -0.25$ while for $\eta=0.5$, $a=0.7$,  $(-3/2) \eta + a /2 = -0.4$. 
Since typically $M \propto L ^{2-2.3}$ in 
these objects, we find that ${M}_{\rm peak} \propto M ^{\simeq 0.1-0.2}$. 
Thus, changing the mass of the clump 
by a factor $10^2$ changes the location of the peak of the IMF by a factor less than $\simeq 2$. 
This partial compensation 
between the comparable increasing and decreasing scale-dependence of the Mach number and the Jeans mass, 
respectively, may be one of the reasons why the peak of the IMF appears to be rather constant over a wide
 range of stellar cluster conditions (see also Elmegreen et al. 2008).

\subsection{Comparison with previous works}
Although based on the same underlying picture of hydrodynamical- or turbulence-driven fragmentation, the
 theory presented in this work is more general than the one proposed by Padoan \& Nordlund (2002). First of all, 
the present theory does not invoke
shock conditions 
whose validity might be questionable, since they assume that the magnetic field is
 perpendicular to the shock. Note that, under this assumption, 
the shock jump conditions imply that $B \propto \rho$, which is not compatible with the observations.
 Second of all,
the present theory does not assume a one-to-one correspondence between the probability distribution of the 
turbulent gas density and that of local Jeans masses, identified as collapsing protostars. Regions
of the gas which will collapse into a gravitationally bound object are properly accounted for, as they
must fulfill the Jeans mass criterion through the proper selection condition for the density threshold.
Regions where density fluctuations fail to fulfill this criterion will not be included into the mass spectrum
distribution of collapsing objects. Therefore, the theory insures a correct counting of the collapsing 
structures. Finally, turbulent or magneto-turbulent support enters explicitly in our theory whereas
it is interesting to note that this is not the case in the Padoan \& Nordlund theory. 
Note that their original CMF/IMF (Padoan et al. 1997) closely
resembles our pure thermal case. It is very  likely the reason why, in the pure hydro case, their
high mass tail is dominated by the steep $dN/dM\propto M^{-3}$ power-law, yielding an underestimate
of the number of massive stars relative to low-mass ones.

As in the Padoan \& Nordlund theory, we recover the fact that the CMF/IMF is shaped by the product of two
contributions, namely a power-law form, which dominates at intermediate scales, 
and two exponential cut-offs, one below about the characteristic mass for collapse and one at large scales.
However, we note that this low-mass
cut-off is not the result of turbulence, but is the result of the Gaussian, more precisely lognormal 
distribution of the density field,
and of the threshold condition for collapsing structures. Although turbulence does produce
such a type of distribution, 
other mechanisms, like wave propagation or gravity,  could presumably lead to similar distributions.
Turbulence, however, leads to a significantly smaller value for the low-mass cut-off, 
compared with the thermal case, promoting the formation of low-mass objects, 
in particular brown dwarfs. 
We also note that in the Padoan \& Nordlund theory, the high mass part of the IMF is a pure power law 
whose index is determined by the shock conditions, with $x=3/(6-n)$ \footnote{We recall that $n$ is here the 3D value
of the power spectrum index. It thus corresponds to $n=\beta+2$ in the Padoan \& Nordlund theory, where $x=3/(4-\beta)$.}, whereas our IMF contains two contributions.  A pure power law, of index $x=(n+1)/(2n-4)$,
and a lognormal contribution resulting in an exponential decrease at the injection scale, as
stated by eqs.(\ref{therm_dens}) and~(\ref{big_dens}).
Interestingly, in the Padoan \& Nordlund theory, $x$ increases with $n$ whereas in our case
$x$ {\it decreases} with $n$. This fundamental difference  is due to the fact that in the Padoan \& Nordlund
theory, turbulence is always promoting star formation since turbulent support is not included,
whereas in our case, turbulence
is also supporting the gas against gravitational collapse, leading globally to a negative effect on star
formation, as shown in \S6.3. Testing these qualitatively different predictions on the
$n$-dependence would bring precious information on the very role of turbulence in star formation.
%In particular, in our case we see that for $n \rightarrow 3$, $x \rightarrow 2$ meaning that we recover 
%the purely thermal case for a turbulence with an energy spectrum, $E(k) \proto k ^{-1}$.

Our theory also explains why
 hydrodynamical simulations of turbulence do recover
the correct Salpeter power-law slope at high masses (Tilley \& Pudritz 2004, Bate \& Bonnell 2005).
It is also in agreement with the finding that the CMF/IMF depends on the Mach 
number (Ballesteros-Paredes et al. 2006).

Finally, our approach provides a consistent explanation, within the same general framework, 
for the mass distribution of both self-gravitating structures and CO clumps. 
This fits well the recent results of Dib et al. (2008) where it is shown that clouds defined by 
a low density threshold present a flat mass spectrum of index $\simeq -1.7$, whereas clouds having a higher degree of 
gravitational binding and defined by a higher density 
threshold  have a stiffer mass spectrum.

\subsection{Restriction of the present work}
It is necessary to emphasize various limitations of our present theory. These issues require further 
detailed investigations.
First of all, as known in the cosmological context, the window function used to perform the smoothing at scale 
$R$ has some influence on the results (e.g. Nagashima 2001). 
The same is true for the probability $P(M,M')$ of finding a structure 
of mass $M$ inside a structure of mass $M'$, as demonstrated by Jedamzik (1995), although Jedamzik 
finds a relatively modest  influence of $P(M,M')$ (although see Yano et al. 1996 and Nagashima 2001). 
Second of all,
the turbulent Jeans mass is treated using a mean scale dependence. Ideally, this could be improved by 
considering a velocity distribution  correlated to the density. The same is true for the magnetic field 
distribution. 

At last, another important aspect already mentioned in \S\ref{time dep},
 ignored in this work as in most of the other theories of the IMF, is the time dependence 
issue in the star formation process.
The present theory consists simply in counting the fluctuations of a given distribution, produced by an underlying
density field.
In principle, the density fluctuations should evolve with time and  rejuvenate. In particular, one 
may wonder whether the small scales should not rejuvenate more rapidly than the larger one.    
We think that part of the answer comes from the assumption of ergoditicity, which states that 
spatial averaging should give the same results as time averaging. In particular, the  ratio of 
small to large scale fluctuations should remain the same.  This is, however, obviously not the case for the scales
comparable to the size of the cloud itself. For those, a time-dependent theory seems unavoidable.

In a related way, our approach does not consider any accretion from external sources nor any further fragmentation during the collapse. 
Both could make the IMF different from the CMF, although the magnetic field seems to reduce 
the fragmentation significantly  (Hennebelle \& Teyssier 2008, Machida et al. 2008, Price \& Bate 2008). On the other hand,
the observed strong similarity between the CMF and the {\it system} IMF suggests that this latter is determined essentially by the CMF and that its shape should not be drastically modified by accretion, other than for the matter already included in the core reservoir. As for subfragmentation of the cores into individual objects, it has been shown that
taking into account the mass-dependent multiplicity frequencies observed in the solar neighborhood seems to provide the correct link between the
IMF of unresolved systems and the one of resolved individual objects (Chabrier 2003a, 2003b). We expect the same to be true for our theoretical
CMF/IMF, although identifying the exact physical processes responsible for this subfragmentation remains an open issue.

\section{Conclusion}

In the present paper, we have derived an analytical theory for the stellar initial mass function and for 
the mass function of the CO clumps, based on
an extension of the statistical Press-Schechter formalism derived in cosmology.
Our theory provides a  {\it predictive} theoretical foundation to understand the origin of the stellar IMF, and to
infer its behaviour in various environments.
The theory predicts that the CMF/IMF involves two contributions, namely a power-law tail and an
exponential cut-off below about the mean thermal or turbulent Jeans mass, even in the
absence of turbulence.
Although thermal collapse produces too steep a slope compared with the Salpeter value, this
latter is recovered exactly in the case of supersonic turbulence, for the appropriate
observed or numerically
determined power spectrum index.  This corroborates the general gravo-turbulent picture of star formation, as initiated by Larson (1981) and developed more recently by Padoan \& Nordlund (2002),
where large-scale protostellar clumps which contain several Jeans masses are dominated by
supersonic turbulent motions and will fragment into prestellar cores that produce the final stellar spectrum. The smaller clumps will have subsonic internal velocities, in agreement with the observed Larson's relations in molecular clouds, i.e. will
be supported by thermal motions, leading to a turnover of the CMF/IMF about the characteristic effective Jeans
mass. 
Turbulence favors the formation of both low-mass and high-mass structures but as a whole
it has a negative effect on star formation, decreasing the overall star formation efficiency. We also suggest that
the opposite, comparable scale-dependences of the Mach number and Jeans mass lead to a weak dependence of the location of the
peak of the IMF upon clump masses, providing an appealing explanation for the observed rather universal behaviour of the IMF over a wide range of stellar cluster conditions.

The success of the present theory in reproducing the CMF/IMF inferred from the observations
of prestellar cores strongly suggests that the IMF is determined by the conditions prevailing in the cloud, temperature,
density and scale-dependence of the velocity dispersion (i.e. characteristic velocity power spectrum), and thus is already
imprinted at early stages in the cloud. The universality of the IMF, at least under present-day Galactic conditions,
very likely arises from the universality of the self-similar nature of turbulence and from comparable characteristic cloud
conditions, determined by the same dominant cooling processes. Effects like accretion, ejection, collisions, winds may affect 
to some extent the exact shape of the CMF/IMF, explaining
possible statistical variations, but are unlikely to be dominant mechanisms.

\acknowledgments
We are grateful to Philippe Andr\'e, Sami Dib, Bruce Elmegreen,  Wolfram Schmidt 
and Enrique V\'azquez-Semadeni 
for insightful comments and for a critical reading 
of the manuscript, to
Edouard Audit for stimulating discussions,  to Paolo Padoan and Aake Nordlund for lively e-mail exchanges and to 
the anonymous referee for helpful comments.
 PH thanks Alexei Kritsuk for many stimulating discussions
on turbulence during his stay at the Kavli Institute for Theoretical Physics and for providing 
the power spectrum of $\log \rho$ calculated with his state of the art hydrodynamical simulations.
He is also very grateful to Chris McKee and Shu-ichiro Inutsuka for many discussions on related topics over the years. 
 GC acknowledges
the warm hospitality of the Max Planck Institute for Astrophysics and numerous discussions with various colleagues.
This work was supported by the french "agence nationale pour la recherche (ANR)" within
the 'magnetic protostars and planets (MAPP)' project and by the "Constellation" european network MRTN-CT-2006-035890.
%This research was supported in part by the National Science Foundation under Grant No. PHY05-51164.

\appendix

\section{Mass spectrum of voids}
\label{voids}
 Here, we show that the voids have the same mass spectrum as the structures.
The voids are regions of gas where the density is smaller than 
the average density. The mass contained within structures of mass smaller
than $M_R=\rho_c R^3$ is given by
\begin{eqnarray}
M_{\rm tot}(R) = L_i^3 \int _{-\infty} ^ {\delta_c} \bar{\rho} \exp(\delta) \,  {\cal P}_R(\delta)\,  d\delta.
\label{voids_eq}
\end{eqnarray}
while eq.(\ref{droit1}) remains applicable. Thus 
\begin{eqnarray}
{\cal N}(M_R) = -{ 2 \bar{\rho}  \over  M_R  } {d R \over dM_R }  { 1 \over \sqrt{2 \pi} \sigma^2} {d \sigma \over dR}
\int ^{\delta_c} _ {-\infty} A(\delta,R)\, d\delta.
\label{form_voids}
\end{eqnarray}
From eq.(\ref{form_A}), we see that
\begin{eqnarray}
\int ^\infty _{-\infty} A(\delta,R) d \delta = 0.
\end{eqnarray}
This implies that
\begin{eqnarray}
\int ^{\delta_c} _{-\infty} A(\delta,R) d \delta = -
\int ^\infty _{\delta_c} A(\delta,R) d \delta .
\end{eqnarray}
Therefore, the mass spectrum of voids is given by
\begin{eqnarray}
{\cal N}(M) = { \bar{\rho}  \over M ^2 } \left( { M \over M_{0}} \right)^{{n'-3\over3} }
\left( {\bar{\rho}  \over \rho_c } \right)^{{n'-3\over3}} 
{  (n'-3) \sigma_0^2 \over 3 \sqrt{2 \pi}  \sigma ^3  } \times 
%\nonumber
\left( -\delta_c - {\sigma^2 \over 2} \right) 
\exp \left(  -{(\delta_c-{\sigma^2\over 2} )^2   \over 2 \sigma^2}  \right).  
\label{n_voids}
\end{eqnarray}

\bigskip

\section{The $d{\cal P}_R/dR$ term}
\label{second_term}
Here, we investigate the role played by the second term which appears in eq.(\ref{n_general}).
From eqs.~(\ref{form_n}),~(\ref{form_A}),~(\ref{crit_ytot}), 
~(\ref{crit_Mtot}) and eq.(\ref{n_general}), eq.(\ref{big_dens}) should be:
\begin{eqnarray}
{\cal N}(\widetilde{M} ) &\simeq& 2\, {\cal N}_0 \, 
{ 1 \over \widetilde{R}^3}  {1 \over 1 + (2 \eta + 1) {\cal M}^2_* \widetilde{R}^{2 \eta} } \nonumber \\
&\times& \left(
{ 1 + (1 - \eta){\cal M}^2_* \widetilde{R}^{2 \eta} \over
 \, (1 + {\cal M}^2_* \widetilde{R}^{2 \eta})^{3/2}} \right. - \left.
{ \delta_R^c + \sigma^2 /2  \over
 \, (1 + {\cal M}^2_* \widetilde{R}^{2 \eta})^{1/2}} {n - 3 \over 2}  {\sigma_0^2 \over \sigma^2}
\left( {\widetilde{R}  \over \widetilde{L}_i}  \right) ^{n-3} 
\right) \nonumber
\\ &\times& 
\exp \left( - { \left[ \ln \left( { \widetilde{M}  /
\widetilde{R}^3 }  \right) \right]^2 \over  2 \sigma^2 } \right)
\times {\exp( -\sigma^2/8 ) \over \sqrt{2 \pi} \sigma } 
\label{big_dens2}
\end{eqnarray}
where $\widetilde{L}_i= L_i / \lambda_J^0$ and where, for simplicity, we have assumed 
 that $n=n'$. 

We thus see that the second term becomes comparable to the first one only when 
$ (\widetilde{R}  / \widetilde{L}_i)^{n-3} \times \sigma_0^2 / \sigma^2(\widetilde{R}) \simeq 1$,
which happens when $R \simeq L_i$, {\it i.e.} for structures whose size is comparable to the injection 
scale, expected to be the size of the whole system.

\section{On the $P(M,M')=1/2$ result}
\label{p=1/2}
Here, for completness, we essentially reproduce the Yano et al. (1996) result which shows that, for a 
window function sharply truncated in the k-space, one has $P(M,M')=1/2$.

Let us consider a subregion of the flow which, smoothed at scale $R'$, has a density 
equal to $\rho_c$. Let $M' \simeq \rho_c (R') ^3$. 
The probability that a region of size $R<R'$ has a density larger 
than $ \rho_c$ and thus contains a mass larger than $M_R^c$, is given by:
\begin{eqnarray}
P(M_R^c,M') = \int ^\infty _{\delta_c} {1 \over \sqrt{2 \pi}} {1 \over \sigma_{\rm sub}}
\exp \left( - {1 \over 2}\left({ \delta - \delta_c \over \sigma_{\rm sub} } \right) ^2 \right) d \delta,
\end{eqnarray}
where $\sigma_{\rm sub} = \sigma^2(R') - \sigma^2(R)$. 

Indeed, this shows that $P(M_R^c,M') = 1/2$.

\section{On the $P(R,M)=1$ assumption}
\label{justif_P=1}
Here we give  justifications of the $P(R,M)=1$ assumption done in the paper. 
Let us consider a spherical cloud with a density profile $\rho \propto r ^{-a} $.
The cloud mass thus grows with $r$ as $M(r) \propto r ^{3-a}$. 

The thermal Jeans mass, $M_J$, is proportional to $1/\sqrt{\rho}$, and thus
  $M_J \propto r ^{a/2}$. This implies that 
\begin{eqnarray} 
{ M_J(r) \over M(r) } \propto r ^{ (3/2) a   -3 }.
\end{eqnarray} 
Therefore, if $a<2$, the Jeans mass grows with $r$ less rapidly than  $M(r)$ and 
a Jeans unstable mass is thus embedded into a larger more gravitationally unstable cloud. This is 
precisely the assumption $P(R,M)=1$. 

In case of turbulent support, $M_J (r) \propto (C_{s, {\rm eff}})^3 / \sqrt{\rho}$, and 
$M_J(r) \propto r ^{3 \eta + a/2}$. Thus, the Jeans mass grows less rapidly than $M(r)$
if $3 \eta + (3 /2) a - 3 < 0$, leading to $a < 2 - 2 \eta$. With $\eta \simeq 0.4-0.5$, this yields
$a < 1-1.2$.

Since it seems reasonable to assume that most cloud density profiles should not be much stiffer than $1/ r^{(1-2)} $, 
we conclude that our assumption $P(R,M)=1$ is realistic.

%\newpage

%\begin{figure}[p]
%\center{\includegraphics[angle=0,width=6in]{fig_compar2.ps}} %{imf_mach.ps}}
%%\center{\includegraphics[angle=0,width=6in]{Figure1cd.pdf}}
%\caption{Mass spectrum for ${\cal M}_*=2.5$, and ${\cal M}=5$ (upper curves) and ${\cal M}=2$
%(lower curves). Dash-line: pure thermal case; dot-line: pure turbulent case; solid line: general case.
%Long-dash line : Salpeter power-law.}
%\label{fig_compar}
%\end{figure}

%\newpage

%\begin{figure}[p]
%\center{\includegraphics[angle=0,width=6in]{fig_Mach.ps}} %{imf_mach_et.ps}}
%\caption{Mass spectrum for Mach values ${\cal M}=3$, 5, 10 and 15, from bottom to top, for
%${\cal M}_*=2.5$. Dotted lines : ${\cal M}=5$, ${\cal M}_*=0.5$ (lower curve) and ${\cal M}=5$, ${\cal M}_*=4.0$ (upper curve).}
%\label{fig_Mach}
%\end{figure}

%\clearpage

\end{document}